\documentclass[aps,prd,nofootinbib,superscriptaddress,twocolumn]{revtex4}

\usepackage{slashed}
\usepackage{hyperref}
\usepackage{amsmath,amsfonts,amssymb}
\usepackage{booktabs}
\usepackage{comment}
\usepackage{siunitx}
\usepackage{xcolor}
\usepackage{verbatim}

\usepackage{comment}
\usepackage{tikz}
\usetikzlibrary{decorations.pathmorphing,decorations.markings}

\tikzset{
  psi/.style={
    decoration={
      markings,
      mark=at position 0.6 with {\arrow{>}}
    },
    postaction={decorate},
    double,
    double distance=1pt
  },
  psiNoArrow/.style={
    decoration={
      markings,
      mark=at position 0.6 with 
    },
    postaction={decorate},
    double,
    double distance=1pt
  },
  nucleon/.style={
    decoration={
      markings,
      mark=at position 0.6 with {\arrow{>}}
    },
    postaction={decorate}
  },
  external/.style={},  
  gluon/.style={
  decorate, draw=black, 
  decoration={coil,amplitude=4pt, segment length=5pt}
  },
  particle/.style={draw=black, postaction={decorate}, decoration={markings,mark=at position .5 with {\arrow[draw=black]{>}}}},
 photon/.style={decorate, decoration={snake,amplitude=2pt, segment length=5pt}, draw=black}
}

\newcommand{\deltaLPiNRdagger}{%
  \begin{tikzpicture}[baseline=-3pt]
    \draw[dashed] (-0.42, 0.42) -- (0, 0);
    \draw[particle] (-0.42, -0.42) -- (0.0, 0.0);
    \draw[psi] (0.42, 0) -- (0.0, 0);

    \node at (-0.54, 0.64) [external]{$\pi(p_1)$};
    \node at (-0.54, -0.54) [external]{$N^{\dagger d}_{R}(p_2)$};
    \node at (0.92, 0) [external]{$\Delta_{L abc}$};
    
  \end{tikzpicture}
}

\newcommand{\deltaRPiNLdagger}{%
  \begin{tikzpicture}[baseline=-3pt]
    \draw[dashed] (-0.42, 0.42) -- (0, 0);
    \draw[particle] (0.0, 0.0) -- (-0.42, -0.42);
    \draw[psi] (0.0, 0) -- (0.42, 0);

    \node at (-0.54, 0.64) [external]{$\pi(p_1)$};
    \node at (-0.54, -0.54) [external]{$N^{\dagger }_{L\dot d}(p_2)$};
    \node at (0.92, 0) [external]{$\Delta_{R}^{\dot a \dot b \dot c}$};
    
  \end{tikzpicture}
}

\newcommand{\deltaLdaggerPiNR}{%
   \begin{tikzpicture}[baseline=-3pt]
    \draw[dashed] (-0.42, 0.42) -- (0, 0);
    \draw[particle] (0.0, 0.0) -- (-0.42, -0.42);
    \draw[psi] (0.0, 0) -- (0.42, 0);

    \node at (-0.54, 0.64) [external]{$\pi(p_1)$};
    \node at (-0.54, -0.54) [external]{$N^{\dot d}_{R}(p_2)$};
    \node at (0.92, 0) [external]{$\Delta_{L \dot a \dot b \dot c}^\dagger$};
    
  \end{tikzpicture}
}

\newcommand{\deltaRdaggerPiNL}{%
  \begin{tikzpicture}[baseline=-3pt]
    \draw[dashed] (-0.42, 0.42) -- (0, 0);
    \draw[particle] (-0.42, -0.42) -- (0.0, 0.0);
    \draw[psi] (0.42, 0) -- (0.0, 0);

    \node at (-0.54, 0.64) [external]{$\pi(p_1)$};
    \node at (-0.54, -0.54) [external]{$N_{L d}(p_2)$};
    \node at (0.92, 0) [external]{$\Delta_{R}^{\dagger abc}$};
    
  \end{tikzpicture}
}

\newcommand{\deltaLGammaNRdagger}{%
  \begin{tikzpicture}[baseline=-3pt]
    \draw[photon] (-0.42, 0.42) -- (0, 0);
    \draw[particle] (-0.42, -0.42) -- (0.0, 0.0);
    \draw[psi] (0.42, 0) -- (0.0, 0);

    \node at (-0.54, 0.64) [external]{$\gamma^\nu(p_1)$};
    \node at (-0.54, -0.54) [external]{$N^{\dagger d}_{R}(p_2)$};
    \node at (0.92, 0) [external]{$\Delta_{L abc}$};
    
  \end{tikzpicture}
}

\newcommand{\deltaRGammaNLdagger}{%
  \begin{tikzpicture}[baseline=-3pt]
    \draw[photon] (-0.42, 0.42) -- (0, 0);
    \draw[particle] (0.0, 0.0) -- (-0.42, -0.42);
    \draw[psi] (0.0, 0) -- (0.42, 0);

    \node at (-0.54, 0.64) [external]{$\gamma^\nu(p_1)$};
    \node at (-0.54, -0.54) [external]{$N^{\dagger }_{L\dot d}(p_2)$};
    \node at (0.92, 0) [external]{$\Delta_{R}^{\dot a \dot b \dot c}$};
    
  \end{tikzpicture}
}

\newcommand{\deltaLdaggerGammaNR}{%
   \begin{tikzpicture}[baseline=-3pt]
    \draw[photon] (-0.42, 0.42) -- (0, 0);
    \draw[particle] (0.0, 0.0) -- (-0.42, -0.42);
    \draw[psi] (0.0, 0) -- (0.42, 0);

    \node at (-0.54, 0.64) [external]{$\gamma^\nu(p_1)$};
    \node at (-0.54, -0.54) [external]{$N^{\dot d}_{R}(p_2)$};
    \node at (0.92, 0) [external]{$\Delta_{L \dot a \dot b \dot c}^\dagger$};
    
  \end{tikzpicture}
}

\newcommand{\deltaRdaggerGammaNL}{%
  \begin{tikzpicture}[baseline=-3pt]
    \draw[photon] (-0.42, 0.42) -- (0, 0);
    \draw[particle] (-0.42, -0.42) -- (0.0, 0.0);
    \draw[psi] (0.42, 0) -- (0.0, 0);

    \node at (-0.54, 0.64) [external]{$\gamma^\nu(p_1)$};
    \node at (-0.54, -0.54) [external]{$N_{L d}(p_2)$};
    \node at (0.92, 0) [external]{$\Delta_{R}^{\dagger abc}$};
    
  \end{tikzpicture}
}

\newcommand{\piNLtoDeltatopiNLdagger}{%
  \begin{tikzpicture}[baseline=-3pt]
    \draw[dashed] (-0.7, 0.7) -- (0, 0);
    \draw[particle] (-0.7, -0.7) -- (0, 0);
    \draw[psi] (1, 0) -- (0, 0);
    \draw[dashed] (1, 0) -- (1.7, 0.7);
    \draw[particle] (1, 0) -- (1.7, -0.7);

    \node at (-0.9, 0.9) [external]{$\pi$};
    \node at (-0.9, -0.9) [external]{$N_L$};
    \node at (0.5, 0.4) [external]{$\Delta$};
    \node at (1.9, 0.9) [external]{$\pi$};
    \node at (1.9, -0.9) [external]{$N_L^\dagger$};
  \end{tikzpicture}
}

\newcommand{\piNLtoDeltatopiNRdagger}{%
  \begin{tikzpicture}[baseline=-3pt]
    \draw[dashed] (-0.7, 0.7) -- (0, 0);
    \draw[particle] (-0.7, -0.7) -- (0, 0);
    \draw[psi] (0.5, 0) -- (0, 0);
    \draw[psi] (0.5, 0) -- (1, 0);
    \draw[dashed] (1, 0) -- (1.7, 0.7);
    \draw[particle] (1.7, -0.7) -- (1, 0);
    
    \node at (-0.9, 0.9) [external]{$\pi$};
    \node at (-0.9, -0.9) [external]{$N_L$};
    \node at (0.5, 0.4) [external]{$\Delta$};
    \node at (1.9, 0.9) [external]{$\pi$};
    \node at (1.9, -0.9) [external]{$N_R^\dagger$};
  \end{tikzpicture}
}

\newcommand{\piNRtoDeltatopiNRdagger}{%
  \begin{tikzpicture}[baseline=-3pt]
    \draw[dashed] (-0.7, 0.7) -- (0, 0);
    \draw[particle] (0, 0) -- (-0.7, -0.7);
    \draw[psi] (0, 0) -- (1, 0);
    \draw[dashed] (1, 0) -- (1.7, 0.7);
    \draw[particle] (1.7, -0.7) -- (1, 0);
    
    \node at (-0.9, 0.9) [external]{$\pi$};
    \node at (-0.9, -0.9) [external]{$N_R$};
    \node at (0.5, 0.4) [external]{$\Delta$};
    \node at (1.9, 0.9) [external]{$\pi$};
    \node at (1.9, -0.9) [external]{$N_R^\dagger$};
  \end{tikzpicture}
}

\newcommand{\piNRtoDeltatopiNLdagger}{%
  \begin{tikzpicture}[baseline=-3pt]
    \draw[dashed] (-0.7, 0.7) -- (0, 0);
    \draw[particle] (0, 0)  -- (-0.7, -0.7) ;
    \draw[psi] (0, 0) -- (0.5, 0);
    \draw[psi] (1, 0) -- (0.5, 0);
    \draw[dashed] (1, 0) -- (1.7, 0.7);
    \draw[particle] (1, 0) -- (1.7, -0.7);
    
    \node at (-0.9, 0.9) [external]{$\pi$};
    \node at (-0.9, -0.9) [external]{$N_R$};
    \node at (0.5, 0.4) [external]{$\Delta$};
    \node at (1.9, 0.9) [external]{$\pi$};
    \node at (1.9, -0.9) [external]{$N_L^\dagger$};
  \end{tikzpicture}
}

\newcommand{\gammaNLtoDeltatopiNLdagger}{%
  \begin{tikzpicture}[baseline=-3pt]
    \draw[photon] (-0.7, 0.7) -- (0, 0);
    \draw[particle] (-0.7, -0.7) -- (0, 0);
    \draw[psi] (1, 0) -- (0, 0);
    \draw[dashed] (1, 0) -- (1.7, 0.7);
    \draw[particle] (1, 0) -- (1.7, -0.7);

    \node at (-0.9, 0.9) [external]{$\gamma$};
    \node at (-0.9, -0.9) [external]{$N_L$};
    \node at (0.5, 0.4) [external]{$\Delta$};
    \node at (1.9, 0.9) [external]{$\pi$};
    \node at (1.9, -0.9) [external]{$N_L^\dagger$};
  \end{tikzpicture}
}

\newcommand{\gammaNLtoDeltatopiNRdagger}{%
  \begin{tikzpicture}[baseline=-3pt]
    \draw[photon] (-0.7, 0.7) -- (0, 0);
    \draw[particle] (-0.7, -0.7) -- (0, 0);
    \draw[psi] (0.5, 0) -- (0, 0);
    \draw[psi] (0.5, 0) -- (1, 0);
    \draw[dashed] (1, 0) -- (1.7, 0.7);
    \draw[particle] (1.7, -0.7) -- (1, 0);
    
    \node at (-0.9, 0.9) [external]{$\gamma$};
    \node at (-0.9, -0.9) [external]{$N_L$};
    \node at (0.5, 0.4) [external]{$\Delta$};
    \node at (1.9, 0.9) [external]{$\pi$};
    \node at (1.9, -0.9) [external]{$N_R^\dagger$};
  \end{tikzpicture}
}

\newcommand{\gammaNRtoDeltatopiNRdagger}{%
  \begin{tikzpicture}[baseline=-3pt]
    \draw[photon] (-0.7, 0.7) -- (0, 0);
    \draw[particle] (0, 0) -- (-0.7, -0.7);
    \draw[psi] (0, 0) -- (1, 0);
    \draw[dashed] (1, 0) -- (1.7, 0.7);
    \draw[particle] (1.7, -0.7) -- (1, 0);
    
    \node at (-0.9, 0.9) [external]{$\gamma$};
    \node at (-0.9, -0.9) [external]{$N_R$};
    \node at (0.5, 0.4) [external]{$\Delta$};
    \node at (1.9, 0.9) [external]{$\pi$};
    \node at (1.9, -0.9) [external]{$N_R^\dagger$};
  \end{tikzpicture}
}

\newcommand{\gammaNRtoDeltatopiNLdagger}{%
  \begin{tikzpicture}[baseline=-3pt]
    \draw[photon] (-0.7, 0.7) -- (0, 0);
    \draw[particle] (0, 0) --  (-0.7, -0.7) ;
    \draw[psi] (0, 0) -- (0.5, 0);
    \draw[psi] (1, 0) -- (0.5, 0);
    \draw[dashed] (1, 0) -- (1.7, 0.7);
    \draw[particle] (1, 0) -- (1.7, -0.7);
    
    \node at (-0.9, 0.9) [external]{$\gamma$};
    \node at (-0.9, -0.9) [external]{$N_R$};
    \node at (0.5, 0.4) [external]{$\Delta$};
    \node at (1.9, 0.9) [external]{$\pi$};
    \node at (1.9, -0.9) [external]{$N_L^\dagger$};
  \end{tikzpicture}
}

\newcommand{\propagatorOne}{%
  \begin{tikzpicture}[baseline=-3pt]
    \draw[psi] (0, 0) -- (1.2, 0);
    
    \draw (0.54, 0.15+0.08) -- (0.738, 0.15+0.08);
    \draw (0.708, 0.18+0.08) -- (0.738, 0.15+0.08);
    \draw (0.708, 0.15+0.08) -- (0.738, 0.15+0.08);
    
    \fill[black] (0,0) circle (0.06cm);
    \fill[black] (1.2,0) circle (0.06cm);
    
    \node at (0, 0.18+0.08) [external]{$(\dot{a})$};
    \node at (1.2, 0.18+0.08) [external]{$(a)$};
    \node at (0.6, 0.4) [external]{$p$};
  \end{tikzpicture}
}

\newcommand{\propagatorTwo}{%
  \begin{tikzpicture}[baseline=-3pt]
    \draw[psi] (1.2, 0) -- (0, 0);
    \draw (0.54, 0.15+0.08) -- (0.738, 0.15+0.08);
    \draw (0.708, 0.18+0.08) -- (0.738, 0.15+0.08);
    \draw (0.708, 0.15+0.08) -- (0.738, 0.15+0.08);
    
    \fill[black] (0,0) circle (0.06cm);
    \fill[black] (1.2,0) circle (0.06cm);
    
    \node at (0, 0.18+0.08) [external]{$(a)$};
    \node at (1.2, 0.18+0.08) [external]{$(\dot{a})$};
    \node at (0.6, 0.4) [external]{$p$};
  \end{tikzpicture}
}

\newcommand{\propagatorThree}{%
  \begin{tikzpicture}[baseline=-3pt]
    \draw[psi] (0.6, 0) -- (0, 0);
    \draw[psi] (0.6, 0) -- (1.2, 0);
    
    \fill[black] (0,0) circle (0.06cm);
    \fill[black] (1.2,0) circle (0.06cm);
    
    \node at (0, 0.18+0.08) [external]{$(b)$};
    \node at (1.2, 0.18+0.08) [external]{$(a)$};
  \end{tikzpicture}
}

\newcommand{\propagatorFour}{%
  \begin{tikzpicture}[baseline=-3pt]
    \draw[psi] (0, 0) -- (0.6, 0);
    \draw[psi] (1.2, 0) -- (0.6, 0);
    
    \fill[black] (0,0) circle (0.06cm);
    \fill[black] (1.2,0) circle (0.06cm);
    
    \node at (0, 0.18+0.08) [external]{$(\dot{b})$};
    \node at (1.2, 0.18+0.08) [external]{$(\dot{a})$};
  \end{tikzpicture}
}

\newcommand{\incomingLeft}{%
  \begin{tikzpicture}[baseline=-3pt]
    \draw[psi] (0, 0) -- (1.2, 0);
    \draw (0.54, 0.15+0.08) -- (0.738, 0.15+0.08);
    \draw (0.708, 0.18+0.08) -- (0.738, 0.15+0.08);
    \draw (0.708, 0.12+0.08) -- (0.738, 0.15+0.08);
    
    \fill[black] (1.2,0) circle (0.06cm);
   
    \node at (1.2, 0.18+0.08) [external]{$(a)$};
    \node at (0.6, 0.3+0.08) [external]{$p,\sigma$};
  \end{tikzpicture}
}

\newcommand{\incomingRight}{%
  \begin{tikzpicture}[baseline=-3pt]
    \draw[psi] (1.2, 0) -- (0, 0);
    \draw (0.54, 0.15+0.08) -- (0.738, 0.15+0.08);
    \draw (0.708, 0.18+0.08) -- (0.738, 0.15+0.08);
    \draw (0.708, 0.12+0.08) -- (0.738, 0.15+0.08);
    
    \fill[black] (1.2,0) circle (0.06cm);
   
    \node at (1.2, 0.18+0.08) [external]{$(\dot{a})$};
    \node at (0.6, 0.3+0.08) [external]{$p,\sigma$};
  \end{tikzpicture}
}

\newcommand{\outgoingLeft}{%
  \begin{tikzpicture}[baseline=-3pt]
    \draw[psi] (0, 0) -- (1.2, 0);
    \draw (0.54, 0.15+0.08) -- (0.738, 0.15+0.08);
    \draw (0.708, 0.18+0.08) -- (0.738, 0.15+0.08);
    \draw (0.708, 0.12+0.08) -- (0.738, 0.15+0.08);
    
    \fill[black] (0,0) circle (0.06cm);
   
    \node at (0, 0.18+0.08) [external]{$(\dot{a})$};
    \node at (0.6, 0.3+0.08) [external]{$p,\sigma$};
  \end{tikzpicture}
}

\newcommand{\outgoingRight}{%
  \begin{tikzpicture}[baseline=-3pt]
    \draw[psi] (1.2, 0) -- (0, 0);
    \draw (0.54, 0.15+0.08) -- (0.738,0.15+0.08);
    \draw (0.708, 0.18+0.08) -- (0.738, 0.15+0.08);
    \draw (0.708, 0.12+0.08) -- (0.738, 0.15+0.08);
    
    \fill[black] (0,0) circle (0.06cm);
   
    \node at (0, 0.18+0.08) [external]{$(a)$};
    \node at (0.6, 0.3+0.08) [external]{$p,\sigma$};
  \end{tikzpicture}
}

\newcommand{\be}{\begin{equation}}
\newcommand{\ee}{\end{equation}}
\newcommand{\bea}{\begin{equation}\begin{aligned}}
\newcommand{\eea}{\end{aligned}\end{equation}}

\newcommand{\hc}{\text{h.c.}}
\newcommand{\ie}{{\it i.e.}}
\newcommand{\eg}{{\it e.g.}}

\newcommand{\gsim}{\lower.7ex\hbox{$\;\stackrel{\textstyle>}{\sim}\;$}}
\newcommand{\lsim}{\lower.7ex\hbox{$\;\stackrel{\textstyle<}{\sim}\;$}}

\newcommand{\nicpb}{Laboratory of High Energy and Computational Physics, NICPB, R\"avala pst. 10, 10143 Tallinn, Estonia}
\newcommand{\ippp}{Institute for Particle Physics Phenomenology, Department of Physics, Durham University, Durham DH1 3LE, United Kingdom}
\newcommand{\capfe}{CAFPE and Departamento de F\'isica Te\'orica y del Cosmos, Universidad de Granada, E-18071 Granada, Spain}

\begin{document}

\title{An effective field theory of the Delta-resonance}

\author{Juan C. Criado}
\affiliation{\ippp}

\author{Abdelhak Djouadi}
\affiliation{\nicpb}
\affiliation{\capfe}

\author{Niko Koivunen}
\affiliation{\nicpb}

\author{Kristjan M\"u\"ursepp}
\affiliation{\nicpb}  
 
\author{Martti Raidal}
\affiliation{\nicpb}

\author{Hardi Veerm\"ae}
\affiliation{\nicpb}

\begin{abstract}
We present an effective field theory of the $\Delta$-resonance as an interacting Weinberg's $(3/2,0)\oplus (0,3/2)$ field in the multi-spinor formalism. We derive its interactions with nucleons $N$, pions $\pi$ and photons $\gamma$, and compute the $\Delta$-resonance cross-sections in pion-nucleon scattering and pion photo-production. The theory contains only the physical spin-3/2 degrees of freedom. Thus, it is intrinsically consistent at the Hamiltonian level and, unlike the commonly used Rarita-Schwinger framework, does not require any additional {\it ad hoc} manipulation of couplings or propagators. The symmetries of hadronic physics select a unique operator for each coupling $N\pi\Delta$ and $\gamma\pi\Delta$. The proposed framework can be extended to also describe other higher-spin hadronic resonances.
\end{abstract}

\maketitle

\section{Introduction}
The $\Delta$-resonance~\cite{Zyla:2020zbs} is the most important baryon resonance. Its mass is close to the mass of the nucleon, and it has significant couplings to nucleons, pions and photons. Albeit not elementary, it is the lightest known particle with spin-3/2. Clearly, a systematic study of the properties of $\Delta$-resonance would require a solid theoretical understanding of the physics of massive spin-3/2 fields.

The Rarita-Schwinger (RS) framework~\cite{Rarita:1941mf}, which is most commonly used to describe physical spin-3/2 particles, including the $\Delta$-resonance, describes these fields as vector-spinors that contain additional unphysical spin-1/2 degrees of freedom (d.o.f.). The free theory for the RS field is fully consistent as the unphysical spin-1/2 d.o.f. is eliminated due to a local symmetry of the free RS Lagrangian. However, the unphysical spin-1/2 d.o.f. can be excited in interacting theories. This leads to theoretical inconsistencies and phenomenological disasters~\cite{Johnson:1960vt,Velo:1970ur}. The difficulties in formulating the $\Delta$-resonance interactions in a Lagrangian description of the RS field are generic and are summarized, \eg, in Ref.~\cite{Benmerrouche:1989uc}. The presence of unphysical spin-1/2 d.o.f. has often been ignored in the computations of $\Delta$-resonance processes and the inconsistent interactions have been thus used~\cite{Peccei:1969sb, Peccei:1969zq, Olsson:1976st, Nozawa:1989pu, Lee:1991pp, Davidson:1991xz, Garcilazo:1993av, Vanderhaeghen:1995fe, Pascalutsa:1995vx, David:1995pi, Scholten:1996mw, Feuster:1996ww, Pascalutsa:1997md, Korchin:1998ff, Mota:1999qp, Lahiff:1999ur}.

The most common approach to avoid these problems has been to construct the RS field interactions so that the unphysical d.o.f. are not involved in physical processes.
This is accomplished by demanding that the interaction terms of the RS field manifest the same symmetries as the free Lagrangian. 
The only known consistent way to remove the unphysical spin-1/2 d.o.f. is to embed the RS theory into a locally supersymmetric set-up \ie, supergravity~\cite{Freedman:1976xh, Deser:1976eh, Freedman:1976py,VanNieuwenhuizen:1981ae}. In supergravity, the aforementioned local symmetry of the free RS Lagrangian is the local supersymmetry, which allows for consistent interactions for the spin-3/2 gravitino~\cite{Grisaru:1977kk,Grisaru:1976vm}, the superpartner of the spin-2 graviton. Certainly, the $\Delta$-resonance cannot be identified with the universally coupled fermionic quanta of gravity.

There have been several attempts to find consistent interactions of spin-3/2 fields in flat space-time. Early steps in this direction were taken in Ref.~\cite{Hagen:1982ez}. The first consistent non-supersymmetric interactions for spin-3/2 fields were presented in Ref.~\cite{Pascalutsa:1998pw, Pascalutsa:1999zz, Pascalutsa:2000kd} and applied to the $\Delta$-resonance\footnote{It was argued that inconsistent interactions linear in the RS field may be acceptable in the context of chiral perturbation theory because the unphysical spin-1/2 d.o.f. can be absorbed into pion-nucleon contact terms~\cite{Pascalutsa:2000kd}.}. However, there is no well-established theory behind these procedures. Instead, the consistent interaction terms are constructed using {\it ad hoc} methods which work for some parameter values but fail for others~\cite{Badagnani:2017una, Badagnani:2015rwj, Badagnani:2015pfa}. Thus, it is unclear whether a potentially inconsistent non-supersymmetric RS starting point can lead to a consistent fundamental theory of massive spin-3/2 fields. 

Covariant field theoretical formulation of spin-3/2 particle interactions is sorely needed for proper description of hadronic resonances, such as in chiral perturbation theory~\cite{Bernard:1994gm, Bernard:1995dp, Hemmert:1996xg, Tang:1996sq, Hemmert:1997ye, Fettes:1998ud, Hacker:2005fh, Wies:2006rv, Pascalutsa:2006up, Mai:2012wy, Chen:2012nx, Hilt:2013uf, Siemens:2016hdi, Yao:2016vbz, Blin:2016itn, Navarro:2019iqj}, covariant isobar models~\cite{Mart:2015jof, Mart:2019mtq, Clymton:2021wof}, coupled channels models~\cite{Anisovich:2009zy, Anisovich:2011fc, Kamano:2013iva, Ronchen:2014cna} and when studying nucleon scattering either non-relativistically~\cite{Sekihara:2016xnq, Sekihara:2021eah} or relativistically~\cite{Pearce:1990uj, Lahiff:1999ur, Gross:1992tj, Sato:1996gk, Feuster:1997pq,Feuster:1998cj, Pascalutsa:1997md, Pascalutsa:2000bs}.
The presence of the $\Delta$-resonance inside compact stars has been studied very recently \cite{Thapa:2021kfo, Raduta:2021xiz, Li:2020ias} and pion-nucleon scattering has also been recently considered in lattice QCD~\cite{Silvi:2021uya}.

To overcome the difficulties associated with higher-spin fields, Weinberg suggested employing these fields in the Lorentz representations $(j,0)\oplus (0,j)$~\cite{Weinberg:1964cn}, where the spin $j$ is arbitrary. These representations contain only the physical higher-spin d.o.f.. Thus, their interactions are not plagued by the issues related to non-physical components. However, this approach does not admit a Lagrangian description\footnote{There were attempts~\cite{Delgado-Acosta:2013kva,DelgadoAcosta:2015ypa,Mart:2019jtb} to derive higher-derivative Lagrangians for these fields. These second-order Lagrangians contain ghosts and lead to pathological theories~\cite{Criado:2020jkp}.} but allows for a consistent calculation of physical observables using interaction Hamiltonians. Unfortunately, the original formulation of this idea has never been applied to the phenomenology of higher-spin fields.

On the other hand, the effective field theory (EFT) re-formulation of Weinberg's original idea using multi-spinor representations~\cite{Criado:2020jkp} has turned out to be simple and practical, providing a consistent description of generic massive higher-spin particle interactions below some cut-off scale $\Lambda$. This EFT has successfully been used to study generic higher-spin dark matter~\cite{Criado:2020jkp}, production and decays of higher-spin particles at colliders~\cite{Criado:2021itq} and higher-spin induced contributions to the anomalous lepton moments~\cite{Criado:2021qpd}. 

In this note, we formulate a theory of the $\Delta$-resonance interacting with nucleons, pions and photons, and compute the observed physical processes mediated by the $\Delta$-resonance. The theory is based on the multi-spinor EFT formalism applied to Weinberg's representation $(3/2,0)\oplus (0,3/2)$. We treat the $\Delta$-resonance as a generic massive spin-3/2 field without specifying how it arises from QCD and hadronic physics. As expected, in the multi-spinor formalism, the $\Delta$ interactions with nucleons and pions, $N\pi\Delta,$ appear already at the level of unbroken isospin symmetry. However, $\Delta$ interactions with photons, $\gamma\pi\Delta$, require a minimal breaking of this symmetry, as the $\Delta$ quadruplet components have different electric charges. After applying the hadronic symmetries respected by the pions, nucleons and photons to the interaction Hamiltonian, we find that, in contrast to the RS formulation, each relevant interaction is controlled by a single coupling constant.

We use the interacting theory to compute the cross-sections of physical processes $\pi N\to \pi N$ and $\gamma N\to \pi N$ mediated by the $\Delta$-resonance. To do so, we conventionally cut off the nucleon momenta from above by applying a nuclear form-factor such as the one of Ref.~\cite{Haberzettl:1998aqi}. To compare our results with the existing ones for the RS field, we re-compute the corresponding cross-sections using the consistent RS interactions of Refs.~\cite{Pascalutsa:1998pw, Pascalutsa:1999zz, Pascalutsa:2000kd}. 

We find that both frameworks reproduce the observed resonant behavior of the $\Delta$-resonance. However, the theoretical foundations of the two approaches are different. The proposed theory can be straightforwardly used for computations of physical processes using simple Feynman rules (that we give in the Appendix). If needed, other interactions can be easily included. Unlike in the RS approach, no additional {\it ad hoc} manipulation of the interactions is needed. Thus the theory works like any other EFT in particle physics, representing a consistent EFT of the $\Delta$-resonance. This is exactly what has been missing so far in the attempts to describe the $\Delta$-resonance. It can be generalized to other higher-spin resonances as well\footnote{For a review of baryon spectroscopy see, \eg, Ref.~\cite{Klempt:2009pi} and for a review on EFTs in nuclear interactions see Ref.~\cite{Epelbaum:2008ga}.}.

This paper is structured as follows. In Section~\ref{sec:theory}, we outline the theoretical framework and derive the interactions for the $\Delta$-resonance. In Section~\ref{sec:our}, we compute the cross-sections for $\Delta$-mediated $\pi N\to \pi N$ and $\gamma N\to \pi N$ scattering and compare them to the cross sections in the RS framework. We conclude in Section~\ref{sec:concl}. A brief summary of the multi-spinor formalism and the Feynman rules are gathered in the Appendix. 

\section{Theoretical framework}
\label{sec:theory}

We describe the system of photons, pions, nucleons and $\Delta$-baryons at low energies through a Lorentz-invariant EFT in which the $\Delta$ is a field in the $(3/2, 0) \oplus (0, 3/2)$ representation of the Lorentz group, using the multi-spinor formalism introduced in Refs.~\cite{Criado:2020jkp, Criado:2021itq, Criado:2021qpd}. From here on, we will refer to this description of a spin-3/2 field as the multi-spinor framework (MSF).\footnote{We acknowledge that this terminology can be mildly misleading since describing higher-spin fields with multi-spinors is not exclusive to the $(j, 0) \oplus (0, j)$ representation. For example, the RS field, which is in the $(1, 1/2) \oplus (1/2, 1)$ representation, can be represented as a multi-spinor object with one dotted and two symmetric undotted spinor indices instead of the customary vector-spinor, \ie, a four-spinor with one Lorentz index.}

\begin{table}[t]
   \centering
    \begin{tabular}{ccccc}
        \toprule
        \hline
        Field & Lorentz irrep & Isospin & Hypercharge & Dimension
         \\
         \hline
        \midrule
        $\pi$ & $\left(0, 0\right)$ & 1 & 0 & 1 \\
        $N_L$ & $\left(1/2, 0\right)$ & 1/2 & 1/2 & 3/2 \\
        $N_R$ & $\left(0, 1/2\right)$ & 1/2 & 1/2 & 3/2 \\
        $\Delta_L$ & $\left(3/2, 0\right)$ & 3/2 & 1/2 & 5/2 \\
        $\Delta_R$ & $\left(0, 3/2\right)$ & 3/2 & 1/2 & 5/2 \\
        $F_L$ & $\left(1, 0\right)$ & 0 & 0 & 2 \\
        $F_R$ & $\left(0, 1\right)$ & 0 & 0 & 2 \\
        \hline
        \bottomrule
    \end{tabular}
    \caption{Fields which appear in the isospin conserving interactions, classified by irreps of the Lorentz and isospin groups.}
    \label{tab:fields}
\end{table}

The internal global symmetry group of the current model is ${\rm SU(2) \times U(1)_Y}$, corresponding to the nuclear isospin $T$ and hypercharge $Y$.
This symmetry is approximate, as it is broken explicitly by gauging the electromagnetic ${\rm U(1)_Q}$ subgroup. 
In Table~\ref{tab:fields} we show the relevant fields, organized into irreps of the Lorentz and internal symmetry groups. The fields $N_{L,R}$ and $\Delta_{L,R}$ correspond to the left- and right-handed components of the nucleon and $\Delta$-baryon multiplets, respectively. $F_{L,R}$ are the left- and right-handed parts of the electromagnetic field strength tensor $(F_L)_{ab} = \sigma^{\mu\nu}_{ab} F_{\mu\nu}$ and $(F_R)_{\dot{a}\dot{b}} = \bar{\sigma}^{\mu\nu}_{\dot{a}\dot{b}} F_{\mu\nu}$, where $ \sigma^{\mu\nu}_{ab} \equiv \frac{i}{4}\left(\sigma^{\mu}_{a\dot a}\bar{\sigma}^{\nu \dot a }_{~~b} - \sigma^{\nu}_{a\dot a}\bar{\sigma}^{\mu \dot a }_{~~b}\right)$ and $\bar\sigma^{\mu\nu}_{\dot a \dot b} \equiv \frac{i}{4}\left(\bar\sigma^{\mu a}_{\dot a}\sigma^{\nu }_{a\dot b} - \bar\sigma^{\nu a}_{\dot a}\sigma^{\mu }_{a\dot b}\right)$. 
We will focus on describing the tree level $\Delta$-mediated $\pi N \to \pi N$ and $\gamma N \to \pi N$ scattering. Since electromagnetic gauge interactions are not involved in these processes, we will not consider them. In the limit of a vanishing electromagnetic coupling constant, ${\rm U(1)_Q}$ acts only on the photon field, and the ${\rm SU(2) \times U(1)_Y}$ symmetry becomes exact.

The relevant scales in this EFT, apart from the masses of the particles, are the pion decay constant $f_{\pi} \sim \SI{100}{MeV}$, and the chiral symmetry breaking scale $\Lambda \simeq 4\pi f_{\pi} \sim \SI{1}{GeV}$. Wilson coefficients of higher-dimensional operators in the Lagrangian are suppressed by products of these two scales. The series in $\pi/f_\pi$ can be resummed using chiral perturbation theory~\cite{Weinberg:1978kz, Gasser:1983yg, Gasser:1984gg} by embedding the pion fields into a non-linear realization of the QCD chiral symmetry. This generates a tower of interactions with any number of pions at each order in $1/\Lambda$. Since we are interested in interactions with the lowest number of pions, we will not perform this resummation and keep only the first terms. We will write Wilson coefficients of order $1/(f^n \Lambda^m)$ as $c / \Lambda^{n + m}$, so the adimensional coefficients $c \sim (4\pi)^n$ may naturally contain factors of $4\pi$.

The lowest-dimensional interaction operators constructed out of the fields in Table~\ref{tab:fields} containing at least one $\Delta$-field have dimension-7. They are given by the following interaction Hamiltonian
\bea
    -\mathcal{H}_{\Delta}
    \!=\!\frac{1}{\Lambda^3} &\Big[
        c_{\pi L} \;
        \partial^a_{\dot{b}} (N_R)^{\dagger\,b}
        \partial^{\dot{b}c} \pi_A T_A (\Delta_L)_{abc}
        \\
        &\;+ c_{F L} \; 
        (F_L)^{ab} (N_R)^{\dagger\,c} \pi_A T_A
        (\Delta_L)_{abc}
        \\
        &\;+ c_{\pi R} \;
        (\Delta_R)^{\dagger}_{abc}
        T_A \partial^a_{\dot{b}} (N_L)^b \partial^{\dot{b} c} \pi_A
        \\
        &\;+ c_{F R} \;
        (\Delta_R)^{\dagger}_{abc}
        (F_L)^{ab} T_A (N_L)^{c} \pi_A
        \\
        &\;+ d^{(1)}_\pi \;
        (\Delta_R)^{\dagger\,abc} T_A i \sigma_B T_A (\Delta_L)_{abc}  
        \pi^B
        \\
        &\;+ d^{(2)}_\pi \;
        (\Delta_R)^{\dagger\,abc} (\Delta_L)_{abc} \pi_A \pi_A
        \\
        &\;+ d^{(3)}_\pi \;
        (\Delta_R)^{\dagger\,abc} T_A T_B (\Delta_L)_{abc} \pi_A \pi_B
        \\
        &\;+ d_F \;
        (\Delta_R)^{\dagger\,abd} (\Delta_L)_{abc} {(F_L)^c}_d
    \Big]\!+\!\text{h.c.}\,,
\eea
where $A$ denotes SU(2) triplet indices, while SU(2) doublet and quadruplet indices are implicit, and $T_A$ are the isospin-1/2 to 3/2 transition matrices given by
\bea
    T_1 &= \frac{1}{\sqrt{6}} \begin{pmatrix}
        -\sqrt{3} & 0 & 1 & 0 \\
        0 & -1 & 0 & \sqrt{3}
    \end{pmatrix},
    \\
    T_2 &= -\frac{i}{\sqrt{6}} \begin{pmatrix}
        \sqrt{3} & 0 & 1 & 0 \\
        0 & 1 & 0 & \sqrt{3}
    \end{pmatrix},
    \\
    T_3 &= \sqrt{\frac{2}{3}} \begin{pmatrix}
        0 & 1 & 0 & 0 \\
        0 & 0 & 1 & 0
    \end{pmatrix}.
\eea
The left- and right-handed nucleons are $N_{La}=(p_{La}~n_{La})^T$ and $N_{R}^{\dot a}=(p_{R}^{\dot a}~n_{R}^{\dot a})^T$, where $p$ is proton and $n$ is neutron. The left- and right-handed $\Delta$-fields are $\Delta_{L abc} = (\Delta^{++}_{Labc} ~ \Delta^{+}_{Labc} ~
\Delta^{0}_{Labc} ~
\Delta^{-}_{Labc})^T$ and $\Delta_{R}^{\dot a \dot b \dot c} = (\Delta^{++\dot a \dot b \dot c}_{R} ~ \Delta^{+\dot a \dot b \dot c}_{R} ~
\Delta^{0\dot a \dot b \dot c}_{R} ~
\Delta^{-\dot a \dot b \dot c}_{R})^T$.
Both $\Delta_{Labc}$ and $\Delta_R^{\dot a\dot b\dot c}$ are totally symmetric in spinor indices.
In the MSF, the $\Delta$-field has mass dimension 5/2 in contrast to the RS framework, where its mass dimension is 3/2. 

To study the $\Delta$-resonance in $\gamma N \to \pi N$ scattering, we need to model the $\Delta N \gamma$ interactions. Since $\Delta$-baryons, nucleons, and photons have isospin $3/2$, $1/2$ and $0$, respectively, these interactions violate isospin symmetry. To construct the $\Delta N \gamma$ interaction terms, we introduce an isospin-$1$, $Y=0$ spurion $S$ as
\bea
    -\mathcal{H}_{\gamma N \Delta}
    &=
    \frac{c_{\gamma L}}{\Lambda^3}
    F^{ab} (N_R)^\dagger T_A (\Delta_L)_{abc} S_A
    \\
    &\phantom{=}
    + \frac{c_{\gamma R}}{\Lambda^3}
    F^{ab} (\Delta_R)^\dagger_{abc} T_A N_L S_A
    + \text{h.c.} \,.
\eea
The $T_3 \neq 0$ components of $S$ are charged, and must vanish for ${\rm U(1)_Q}$ to be preserved, so we will take $S_3 = 1$. Invariance under parity imposes further relations between the Wilson coefficients. Parity transformation acts as $\pi \leftrightarrow -\pi$, $N_L \leftrightarrow N_R$, $\Delta_L \leftrightarrow \Delta_R$, $F_L \leftrightarrow F_R$. Therefore,
\bea
&   c_{\pi L} = c_{\pi R}^*, \quad 
    c_{F L} = c_{F L}^*, \quad
    c_{\gamma L} = c_{\gamma R}^*, \quad    
    \\&\qquad\quad
    d^{(k)}_\pi = d^{(k)*}_\pi, \quad d_F = d_F^*,
    \label{eq:parity}
\eea
implying that there is a single $\Delta N \pi$ interaction. With these relations, the interaction terms can be expressed using the parity eigenstates $\pi$, $N = N_L \oplus N_R$, $\Delta = \Delta_L \oplus \Delta_R$, $F$ and $\widetilde{F}$. However, we will not follow this path since these fields do not transform as irreps of the Lorentz group and do not fit well into the MSF. 

To conclude, the relations \eqref{eq:parity} imply that $\pi N \to \pi N$ and $\gamma N \to \pi N$ scatterings are controlled by two complex parameters $c_\pi$ and $c_\gamma$ with the interactions given by
\bea
    -\mathcal{H}_{\pi N \Delta}
    &=
    \frac{c_\pi}{\Lambda^3} \Big[
        \partial^a_{\dot{b}} (N_R)^{\dagger\,a}
        \partial^{\dot{b}c} \pi_A T_A (\Delta_L)_{abc}
    \\
    &\phantom{= \frac{c_\pi}{\Lambda^3} \Big[}
        + \partial_{\dot{a}b} (N_L)^\dagger_{\dot{b}}
        \partial^b_{\dot{c}} \pi_A T_A (\Delta_R)^{\dot{a}\dot{b}\dot{c}}
    \Big]
    + \text{h.c.}\,,
    \\
    -\mathcal{H}_{\gamma N \Delta}
    &=
    \frac{c_{\gamma}}{\Lambda^2} \Big[
        F^{ab} (N_L)^c T_3 (\Delta_L)_{abc}
    \\
    &\phantom{= \frac{c_\gamma}{\Lambda^2} \Big[}
        + F_{\dot{a}\dot{b}} (N_L)_{\dot{c}} T_3
        (\Delta_L)^{\dot{a}\dot{b}\dot{c}}
    \Big]
    + \text{h.c.}\,.
\eea
\section{Results and discussion}
\label{sec:our}

\begin{figure}[t]
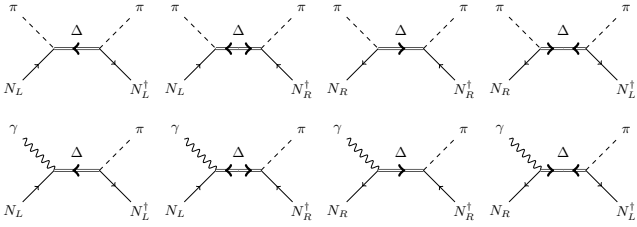

    \centering
    \scalebox{0.6}{\piNLtoDeltatopiNLdagger \piNLtoDeltatopiNRdagger
    \piNRtoDeltatopiNRdagger
    \piNRtoDeltatopiNLdagger}\\[2mm]
    \scalebox{0.6}{\gammaNLtoDeltatopiNLdagger \gammaNLtoDeltatopiNRdagger
\gammaNRtoDeltatopiNRdagger
\gammaNRtoDeltatopiNLdagger}
    \caption{\!Feynman diagrams contributing to the resonant $\pi N \! \to \! \pi N$  (top) and $\gamma N \! \to \! \pi N$ (bottom) processes.}
    \label{piNtopiNdiagrams}
\vspace*{-2mm}
\end{figure}

Using MSF, we compute the cross-sections for $\Delta$-baryon mediated $\pi N \to \pi N$ and $\gamma N \to \pi N$ scattering. For simplicity, we omit the details related to isospin and hence drop the isospin transition matrices from the interaction terms. The $\pi N \Delta$ and $\gamma N\Delta$ interactions then become
\bea\label{eq:L_MS}
    -\mathcal{H}^{\rm MSF}_{\pi N\Delta}  = \frac{1}{\Lambda^3}\Bigg[ 
 & c_L \Big(\partial^a_{~\dot b}N^{\dagger b}_R\Big) \Big(\partial^{\dot b c}\pi\Big)\Big(\Delta_{Labc}\Big)  
 \\
+
& c_L \Big(\partial_{\dot a}^{~ b}N^{\dagger }_{L \dot b}\Big) \Big(\partial_{b  \dot c}\pi\Big)\Big(\Delta_{R}^{\dot a \dot b \dot c}\Big)\Bigg]+ \hc\,,
    \\
    -\mathcal{H}^{\rm MSF}_{\gamma N\Delta} = \frac{1}{\Lambda^2}\Bigg[  
& c_\gamma N^{\dagger a}_R\sigma^{bc}_{\mu\nu}\Delta_{Labc} F^{\mu\nu} 
    \\
+ & c_\gamma N^{\dagger}_{L\dot a}\bar\sigma_{\dot b\dot c\mu\nu}\Delta_{R}^{\dot a\dot b\dot c} F^{\mu\nu}
\Bigg] +\hc \,.
\eea
The resonant Feynman diagrams contributing to the $\pi N\to \pi N$ and $\gamma N\to \pi N$ processes are displayed in the upper and lower parts of Fig.~\ref{piNtopiNdiagrams}, respectively. There are additional diagrams with $u$-channel contributions but they are not shown in this figure.

In the case of $\pi N\to \pi N$ scattering, our result for the cross-section in the MSF is given by
\be\label{eq:sigma_pi_MSF}
    \sigma^{\rm MS}_{\pi N\to \pi N} 
    = \frac{1}{576\pi s}\frac{c_L^4}{\Lambda^{12}}\frac{s^8 P\left(m_\Delta^2/s\right)}{(s-m_\Delta^2)^2+m_\Delta^2 \Gamma^2_\Delta},
\ee
where
\bea
    P(x) 
    = & 1 + 1.1 x + 0.28 x^2 - 4.4 x^3 + 3.5 x^4 - 3.3x^5 \\
    + & 3.6 x^6 - 1.5 x^7 + 0.11 x^8 + 0.033 x^9.
\label{eq:P}
\eea
The coefficients of $P(x)$, which are in general involved functions of the masses, have been computed assuming $m_\pi=0.135$~GeV, $m_N=0.938$~GeV and $m_\Delta=1.232$~GeV for the pion, nucleon and $\Delta$-resonance masses. 

For the $\gamma N\to \pi N$ process, the MSF cross-section has a non-polynomial $s$-dependence and we thus only present its high energy limit given by 
\bea\label{eq:sigma_g_MSF}
    \sigma^{\rm MS}_{\gamma N\to \pi N}\simeq\frac{1}{32\pi s}\frac{c_L^2 c_\gamma^2}{\Lambda^{10}}\frac{ 0.19 s^7 }{(s-m_\Delta^2)^2+m_\Delta^2\Gamma^2_\Delta}.
\eea

\begin{figure*}
    \centering
    \includegraphics[width=0.49\textwidth]{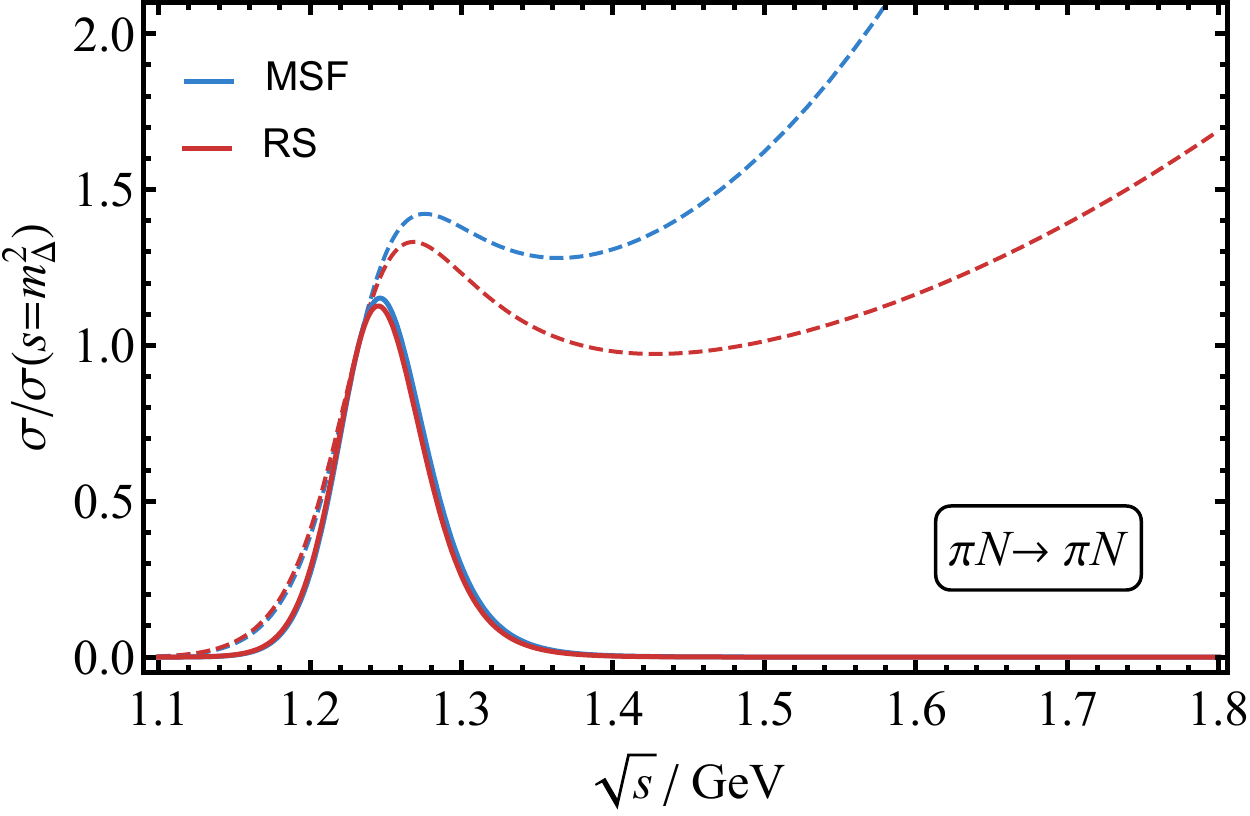}
    \includegraphics[width=0.49\textwidth]{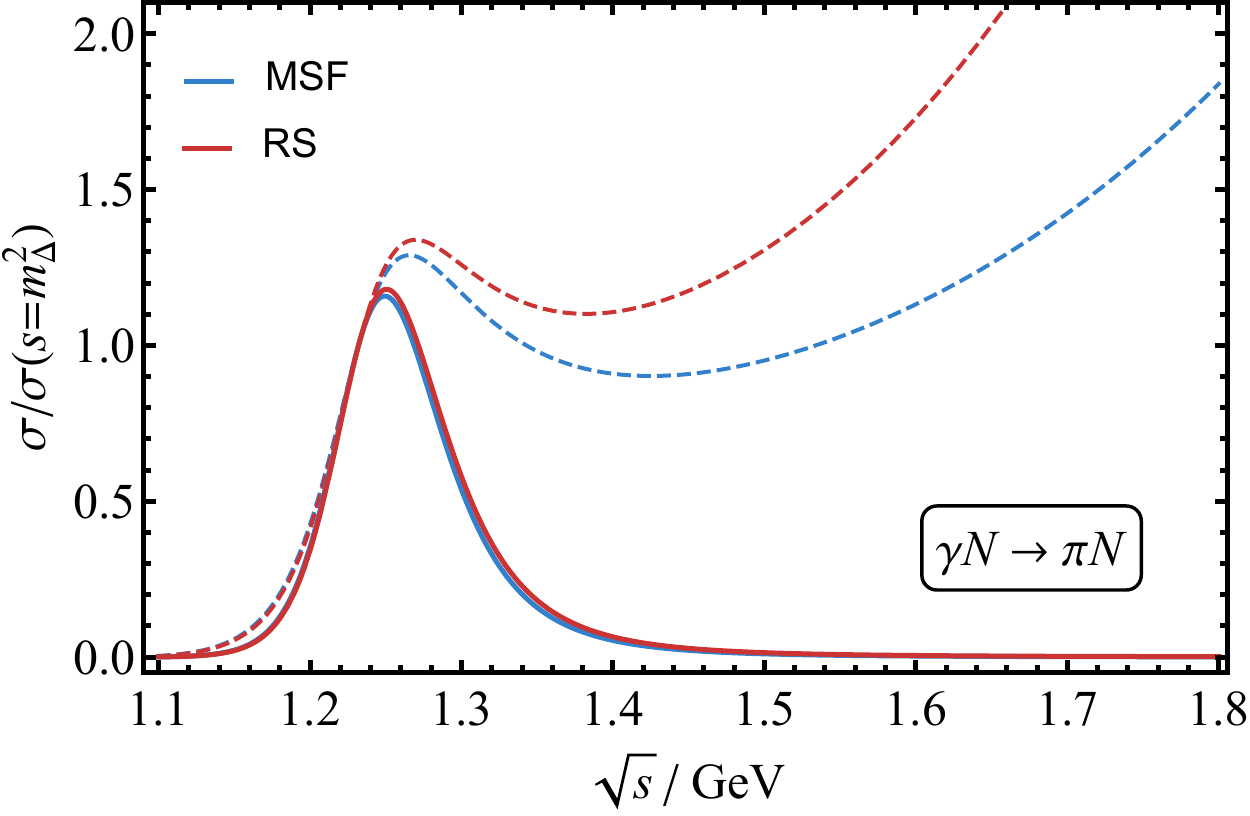}
    \caption{Cross-sections of the $\pi N$ scattering (left) and the pion photoproduction (right) processes in the multi-spinor (blue) and in the Rarita-Schwinger (red) frameworks. The cross-sections are normalized to unity at $s = m_\Delta^2$. The solid lines depict physical cross-sections convoluted over the form-factor \eqref{eq:FF}, while the dashed lines show the behavior of the EFT cross-sections.}
    \label{fig:PiNFull}
\end{figure*}

The cross sections for $\pi N\to \pi N$ and $\gamma N\to \pi N$ scattering in the MSF as functions of the c.m. energy are displayed by the dashed blue lines in the left and right panel of Fig.~\ref{fig:PiNFull}, respectively. The unrestricted growth of these cross-sections beyond the resonance peak is clearly unphysical. However, as hadrons are composite objects, one must use appropriate form-factors to cut off the high momenta in order to capture their experimentally observed behavior. In our study, we consider the form-factor presented in Ref.~\cite{Haberzettl:1998aqi},
\be\label{eq:FF}
    F = \frac{\Lambda^4}{\Lambda^4 + (s-m_{\Delta}^2)^2} , 
\ee
which should be is inserted into each vertex containing only hadrons. The cross-sections for the $\Delta$-mediated $\pi N\to \pi N$ and $\gamma N\to \pi N$ in the MSF by applying the form-factor \eqref{eq:FF} and normalized to unity at $s = m_\Delta^2$, are shown in Fig.~\ref{fig:PiNFull} with blue solid lines. Their behavior is clearly physical, proving that our theory is capable of explaining the hadronic resonances.

The MSF cross-sections should be compared with those obtained in the RS framework. In this case, 
we will consider the $\pi N \Delta$ and $\gamma N\Delta$ interactions that preserve the symmetries of the free RS Lagrangian and thus avoid inconsistencies related to unphysical spin-1/2 d.o.f.~\cite{Pascalutsa:1999zz},
\bea\label{eq:L_RS}
    \mathcal{L}^{\rm RS}_{\pi N\Delta} 
    = f \bar{N} & \gamma_5 \widetilde{G}^{\mu\nu}\partial_\nu \pi+ \hc,
    \\
    \mathcal{L}^{RS}_{\gamma N\Delta} 
    = \bar{N} \Big[ &
    g_1 \widetilde{G}_{\mu\nu}
    +g_2\gamma_5 G_{\mu\nu} + g_3\gamma_\mu\gamma^\rho \widetilde{G}_{\rho\nu}\\
    + & 
    g_4\gamma_5\gamma_\mu\gamma^\rho G_{\rho\nu}
    \Big] F^{\mu\nu}+\hc,
\eea
where $f$, $g_i$, $i=1,...,4$ are dimensionful coupling constants,  $G^{\mu\nu}\equiv\partial^\mu\Delta^\nu-\partial^\nu\Delta^\mu$ is the manifestly invariant RS field tensor and $\widetilde{G}^{\mu\nu} \equiv \frac12 \epsilon^{\mu\nu\rho\sigma} G_{\rho\sigma}$ its dual. Here, the isospin-structure has been omitted. 

The RS framework is an EFT with the $\pi N\Delta$ and $\gamma N\Delta$ interactions given by dimension-6 operators. Notably, the $\gamma N \Delta$ interactions presented in Eq.~\eqref{eq:L_RS} depend on four free coupling constants, whereas the MSF contains a single $\gamma N\Delta$ coupling. The interactions for $\pi N\Delta$ in both frameworks involve two space-time derivatives. On the other hand, the $\gamma N\Delta$ interactions contain two derivatives in the RS but only a single derivative in the MSF. In both approaches, the vertices introduce additional momentum dependence. 

The propagator for the $\Delta$-resonance, however, behaves quite differently. In the MSF (see Appendix~\ref{app:feyn}), it has at most three powers of momentum in the numerator, effectively contributing one power of momentum to the scattering amplitude. The propagator of the RS field, instead, takes the following form\footnote{We have set $A=-1$ as needed for consistent RS interactions~\cite{Pascalutsa:2000kd}.}
\bea
    S^{\mu\nu}(p) 
    & =\frac{\slashed p+ m}{p^2-m^2+i\epsilon}\Big[-\eta^{\mu\nu}+\frac{1}{3}\gamma^\mu\gamma^\nu\\
    & +\frac{1}{3m}(\gamma^\mu p^\nu-\gamma^\nu p^\mu)+\frac{2}{3m^2}p^\mu p^\nu\Big].
\eea
Like in the MSF, the numerator of the RS propagator contains terms that are at most cubic in momentum, but the last two terms decouple when used with consistent interactions, yielding an energy dependence in the ultraviolet regime that is milder than in the MSF. The resulting $\pi N \to \pi N $ cross-section is given by
\be\label{eq:sigma_pi_RS}
    \sigma^{\rm RS}_{\pi N \to \pi N } 
    = \frac{f^4}{576 \pi s}\frac{s^6 \widetilde{P}\left(m_\Delta^2/s\right)}{(s-m_{\Delta}^2)^2 + m_{\Delta}^2\Gamma^2} ,
\ee
where
\bea
    \widetilde{P}(x)~ 
    =~& 1 + 8.2 x -16  x^2 + 4.0  x^3\\
    + & 6.1  x^4 - 3.8  x^5 + 0.51  x^6 + 0.033  x^7.
\eea
with the coefficients evaluated for the same masses as for Eq.~\eqref{eq:P}. In this case, the process $\gamma N \to \pi N$ depends on four coupling constants, which we set equal for simplicity, $g\equiv g_1=g_2=g_3=g_4$. The analytic form of the $\gamma N\to \pi N$ cross-section is again complicated, which is why we only present the high-energy limit of the numerator,
\be
    \sigma^{\rm RS}_{\gamma N\to \pi N} 
    \simeq 
    \frac{f^2 g^2}{32\pi s}\frac{ 0.6 ~s^6 }{(s-m_\Delta^2)^2+m_\Delta^2\Gamma^2_\Delta}.
\ee

We compare our results for the MSF (blue) and the RS framework (red) for the processes $\pi N\to \pi N$ (left panel) and $\gamma N\to \pi N$ (right panel) in Fig.~\ref{fig:PiNFull}. The solid (dashed) lines depict the cross-section with (without) the correction from the form-factor Eq.~\eqref{eq:FF}. Omitting the form-factor, the  $\pi N\to \pi N$ and $\gamma N\to \pi N$ cross-sections in the MSF grow as $\sim s^5$ and $\sim s^4$, respectively. On the other hand, in the RS scenario, both cross-sections scale as $\sim s^3$ at high energies. For the $\pi N\to \pi N$ scattering, this behavior is visible in Fig.~\ref{fig:PiNFull}, while for $\gamma N\to \pi N$ the result in the MSF starts to be dominating at energies higher than those shown in the figure.

All in all, the MSF nicely describes the behavior of the $\Delta$-resonance, as it is evident from Fig.~\ref{fig:PiNFull}. We believe that the same construction could also be used to describe other spin-3/2 resonances. The generalization to spins higher than 3/2 should also be possible\footnote{The RS formalism of Ref.~\cite{Pascalutsa:1998pw, Pascalutsa:1999zz, Pascalutsa:2000kd} was generalized to spins higher than 3/2 in Ref.~\cite{Vrancx:2011qv}.}. It would, in any case, be required to describe all known higher-spin resonances consistently.

\section{Conclusions}
\label{sec:concl}

We have constructed an effective field theory of the spin-3/2 $\Delta$-resonance that describes it as an interacting Weinberg's $(3/2,0)\oplus (0,3/2)$ field in the multi-spinor formalism~\cite{Criado:2020jkp}. In this framework, only the physical degrees of freedom of the $\Delta$-resonance appear. This should be contrasted with the Rarita-Schwinger framework in which the $(1, 1/2) \oplus (1/2, 1)$ field also contains unphysical spin-1/2 components that must be excluded from the physical spectrum using additional symmetries. In the proposed multi-spinor formalism, the interaction terms follow solely from the isospin, Lorentz-, parity- and CP-invariance, and they cannot spoil the counting of degrees of freedom. Therefore, the theory is automatically consistent and can be used for physical computations without any additional assumption, like any other effective field theory in particle physics.

Using the proposed multi-spinor formalism, we derived the most general $\pi N\Delta$ and $\gamma N\Delta$ interaction terms and computed the resulting $\pi N\to \Delta^\ast\to \pi N$ and $\gamma N\to \Delta^\ast \to\pi N$ cross-sections. We compared these cross-sections with the ones computed in the Rarita-Schwinger framework and found that, formally, both approaches can reproduce the observed resonant behavior. 

However, we argue that the multi-spinor framework provides a theoretically consistent and easily implementable effective framework for studying the $\Delta$-resonance. Since this field theory is formulated using physical degrees of freedom only, it provides a promising avenue for extending the model of the $\Delta$-resonance to include new interactions not considered in this work, and for constructing effective theories describing other hadronic higher-spin resonances.

\vspace{5mm}
\noindent \textbf{Acknowledgement.} 
This work was supported by the Estonian Research Council grants MOBTTP135, PRG803, MOBTT5, MOBJD323 and MOBTT86, and by the EU through the European Regional Development Fund CoE program TK133 ``The Dark Side of the Universe." J.C.C. is supported by the STFC under grant ST/P001246/1 and A.D. by the Junta de Andalucia through the Talentia Senior program. 

\clearpage
\appendix

\section{Symmetric multi-spinor formalism}
\label{app:formalism}
\vspace*{-3mm}

The multi-spinor notation, first presented in Ref.~\cite{Criado:2020jkp}, is based on the well established two-component spinor formalism (see Refs.~\cite{Dreiner:2008tw, Borodulin:2017pwh}). Dotted indices ($\dot a,\dot b,\ldots$) and undotted indices ($a,b,\ldots$) transform in the $(1/2, 0)$ and $(0,1/2)$ irrep of the Lorentz group, respectively. The indices are raised and lowered with antisymmetric $\epsilon_{ab}$- and $\epsilon_{\dot a \dot b}$-symbols with $\epsilon_{12} = -\epsilon^{12} = 1$. We utilize the standard convention where dotted (undotted) indices are contracted in ascending (descending) order. The notable exception to this rule is the raising and lowering of indices, \eg~ $t^{a} = \epsilon^{ab}t_{b}$, $t^{\dot a} = t_{\dot a}\epsilon^{\dot a \dot b}$ so that $t^{a} t_{a} = t_{b}\epsilon^{ab}t_{a}$. A pair consisting of a dotted and an undotted spinor index can be converted into a vector index $\mu$, \ie, $p_{a\dot a} = p_{\mu}\sigma^\mu_{a\dot{a}}$ or $p_{\dot a a} = p_{\mu}\bar\sigma^\mu_{\dot{a} a}$ and inversely $p^{\mu} = {\sigma}^{\mu }_{a\dot a}p^{\dot a a}/2$ or $p^{\mu} = \bar{\sigma}^{\mu \dot{a} a}p_{a\dot a}/2$. In $\sigma^\mu$, the $\sigma^0$ is the identity matrix and $\sigma^i$ with $i\!=\!1, 2, 3$ are the Pauli matrices; $\bar\sigma^\mu = (\sigma^{0}, -\sigma^{i})$.

We use the bracket-notation around the index to represent a completely symmetric multi-spinor index.
Objects in the $(j,0)$ irrep are denoted by $t_{(a)} \equiv t_{( a_1 a_2 \ldots a_{2j})}$ and those in the $(0,j)$ irrep by $t_{(a)} \equiv t_{(\dot a_1 \dot a_2 \ldots \dot a_{2j})}$ where all the spinor indices are symmetrized. The objects in $(j,j)$ rep are denoted as $t_{(a)(\dot a)}$. For example, in the case of spin-3/2, the momentum $p_{a\dot{a}}$ corresponds to the multi-spinor object
$$
    p_{(a)(\dot{a})} \equiv \frac{1}{3!} 
    \big[\,
        p_{a_1\dot{a}_1} p_{a_2\dot{a}_2}p_{a_3\dot{a}_3} + \textrm{permutations }
    \big].
$$
In a similar way, the $\epsilon_{ab}$ , $\epsilon^{ab}$ and $\delta^{a}_{b}$ symbols are generalized to $\epsilon_{(a)(b)}$, $\epsilon^{(a)(b)}$ and $\delta^{(a)}_{(b)}$ symbols that can be used to raise and lower symmetric multi-spinor indices.

\vspace*{-3mm}
\section{Feynman rules}
\label{app:feyn}
\vspace*{-3mm}

Here we present the Feynman rules for $\Delta$-resonance propagators and vertices. All vertices are completely symmetric in the spinor indices. Below, $(a)$ and $(\dot a)$ stand for three totally symmetric spinor indices.\bigskip

\noindent {\it $\Delta$-Propagators:}\\[-4mm]

\begin{eqnarray*}
    &\propagatorOne = i\frac{p_{(a)(\dot{a})}}{p^2-m^2} , \
    \propagatorTwo = i\frac{p^{(\dot{a})(a)}}{p^2-m^2} , \\
    &\propagatorThree = i\frac{m^{3}\delta^{(b)}_{(a)}}{p^2-m^2} , \
    \propagatorFour  = i\frac{m^{3}\delta^{(\dot{a})}_{(\dot{b})}}{p^2-m^2}.
\end{eqnarray*}

\noindent {\it External lines:}\\[-4mm]

\begin{eqnarray*}
    \incomingLeft &= u_{(a)}(p,\sigma), \,  
    \incomingRight  = v^{\ast}_{(\dot{a})}(p,\sigma),
\\
    \outgoingLeft & = u^\ast_{(\dot{a})}(p,\sigma), \,
    \outgoingRight  = v_{(a)}(p,\sigma).
\end{eqnarray*}
\vspace*{1mm}

\newpage

\noindent{\it Completeness relations:}\\[-4mm]

\begin{eqnarray*}
    \sum_\sigma u_{(a)}(p,\sigma)  u^\ast_{(\dot{a})}(p,\sigma) =   p_{(a)(\dot{a})},\\
    \sum_\sigma v_{(a)}(p,\sigma) v^\ast_{(\dot{a})}(p,\sigma) =   p_{(a)(\dot{a})},\\
    \sum_\sigma u_{(a)}(p,\sigma) v^{(b)}(p,\sigma) =  m^{3}\delta^{(b)}_{(a)}, \\
    \sum_\sigma {u^\ast}^{(\dot a)}(p,\sigma) {v^\ast}_{(\dot b)}(p,\sigma) =  m^{3}\delta^{(\dot a)}_{(\dot b)} .
\end{eqnarray*}

\noindent {\it Interaction vertices:}\\[-4mm]

\begin{eqnarray*}
&\deltaLPiNRdagger = -i\frac{c_L}{\Lambda^3}\sigma^{\mu a}_{~~ \dot b} \bar\sigma^{\nu\dot b c}~ p_{2\mu} p_{1\nu}~\delta^b_d\\
&\deltaRPiNLdagger = -i\frac{c_L}{\Lambda^3}\bar\sigma^{\mu b}_{\dot a} \sigma^{\nu}_{b \dot c}~ p_{2\mu} p_{1\nu}~\delta^{\dot d}_{\dot b}\\
&\deltaLdaggerPiNR =  i\frac{c_L}{\Lambda^3}\bar\sigma^{\nu \dot c b} \sigma^{\mu \dot a}_{b}~ p_{2\mu} p_{1\nu}~\delta^{\dot b}_{\dot d}\\
&\deltaRdaggerPiNL = i\frac{c_L}{\Lambda^3}\sigma^{\nu}_{c \dot b} \bar\sigma^{\mu\dot b}_{a}~ p_{2\mu} p_{1\nu}~\delta^d_b\\
\\
&\deltaLGammaNRdagger = \frac{2c_\gamma}{\Lambda^2}(\sigma_{\mu} \bar\sigma_{\nu})^{bc}~  p_1^{\mu}~\delta^a_d\\
&\deltaRGammaNLdagger = \frac{2c_\gamma}{\Lambda^2}(\bar\sigma_{\mu} \sigma_{\nu})_{\dot b\dot c}~  p_1^{\mu}~\delta^{\dot d}_{\dot a}\\
&\deltaLdaggerGammaNR =  \frac{2c_\gamma}{\Lambda^2}(\bar\sigma_{\nu} \sigma_{\mu})^{\dot c\dot b}~  p_1^{\mu}~\delta^{\dot a}_{\dot d}\\
&\deltaRdaggerGammaNL = \frac{2c_\gamma}{\Lambda^2}(\sigma_{\nu} \bar\sigma_{\mu})_{cb}~  p_1^{\mu}~\delta^d_a\\
\end{eqnarray*}

\bibliography{NP_resonance}

\begin{thebibliography}{85}
\expandafter\ifx\csname natexlab\endcsname\relax\def\natexlab#1{#1}\fi
\expandafter\ifx\csname bibnamefont\endcsname\relax
  \def\bibnamefont#1{#1}\fi
\expandafter\ifx\csname bibfnamefont\endcsname\relax
  \def\bibfnamefont#1{#1}\fi
\expandafter\ifx\csname citenamefont\endcsname\relax
  \def\citenamefont#1{#1}\fi
\expandafter\ifx\csname url\endcsname\relax
  \def\url#1{\texttt{#1}}\fi
\expandafter\ifx\csname urlprefix\endcsname\relax\def\urlprefix{URL }\fi
\providecommand{\bibinfo}[2]{#2}
\providecommand{\eprint}[2][]{\url{#2}}

\bibitem[{\citenamefont{Zyla et~al.}(2020)}]{Zyla:2020zbs}
\bibinfo{author}{\bibfnamefont{P.~A.} \bibnamefont{Zyla}} \bibnamefont{et~al.}
  (\bibinfo{collaboration}{Particle Data Group}), \bibinfo{journal}{PTEP}
  \textbf{\bibinfo{volume}{2020}}, \bibinfo{pages}{083C01}
  (\bibinfo{year}{2020}).

\bibitem[{\citenamefont{Rarita and Schwinger}(1941)}]{Rarita:1941mf}
\bibinfo{author}{\bibfnamefont{W.}~\bibnamefont{Rarita}} \bibnamefont{and}
  \bibinfo{author}{\bibfnamefont{J.}~\bibnamefont{Schwinger}},
  \bibinfo{journal}{Phys. Rev.} \textbf{\bibinfo{volume}{60}},
  \bibinfo{pages}{61} (\bibinfo{year}{1941}).

\bibitem[{\citenamefont{Johnson and Sudarshan}(1961)}]{Johnson:1960vt}
\bibinfo{author}{\bibfnamefont{K.}~\bibnamefont{Johnson}} \bibnamefont{and}
  \bibinfo{author}{\bibfnamefont{E.}~\bibnamefont{Sudarshan}},
  \bibinfo{journal}{Annals Phys.} \textbf{\bibinfo{volume}{13}},
  \bibinfo{pages}{126} (\bibinfo{year}{1961}).

\bibitem[{\citenamefont{Velo and Zwanziger}(1969)}]{Velo:1970ur}
\bibinfo{author}{\bibfnamefont{G.}~\bibnamefont{Velo}} \bibnamefont{and}
  \bibinfo{author}{\bibfnamefont{D.}~\bibnamefont{Zwanziger}},
  \bibinfo{journal}{Phys. Rev.} \textbf{\bibinfo{volume}{188}},
  \bibinfo{pages}{2218} (\bibinfo{year}{1969}).

\bibitem[{\citenamefont{Benmerrouche et~al.}(1989)\citenamefont{Benmerrouche,
  Davidson, and Mukhopadhyay}}]{Benmerrouche:1989uc}
\bibinfo{author}{\bibfnamefont{M.}~\bibnamefont{Benmerrouche}},
  \bibinfo{author}{\bibfnamefont{R.}~\bibnamefont{Davidson}}, \bibnamefont{and}
  \bibinfo{author}{\bibfnamefont{N.}~\bibnamefont{Mukhopadhyay}},
  \bibinfo{journal}{Phys. Rev. C} \textbf{\bibinfo{volume}{39}},
  \bibinfo{pages}{2339} (\bibinfo{year}{1989}).

\bibitem[{\citenamefont{Peccei}(1969)}]{Peccei:1969sb}
\bibinfo{author}{\bibfnamefont{R.~D.} \bibnamefont{Peccei}},
  \bibinfo{journal}{Phys. Rev.} \textbf{\bibinfo{volume}{181}},
  \bibinfo{pages}{1902} (\bibinfo{year}{1969}).

\bibitem[{\citenamefont{Peccei}(1968)}]{Peccei:1969zq}
\bibinfo{author}{\bibfnamefont{R.}~\bibnamefont{Peccei}},
  \bibinfo{journal}{Phys. Rev.} \textbf{\bibinfo{volume}{176}},
  \bibinfo{pages}{1812} (\bibinfo{year}{1968}).

\bibitem[{\citenamefont{Olsson and Osypowski}(1978)}]{Olsson:1976st}
\bibinfo{author}{\bibfnamefont{M.~G.} \bibnamefont{Olsson}} \bibnamefont{and}
  \bibinfo{author}{\bibfnamefont{E.~T.} \bibnamefont{Osypowski}},
  \bibinfo{journal}{Phys. Rev. D} \textbf{\bibinfo{volume}{17}},
  \bibinfo{pages}{174} (\bibinfo{year}{1978}).

\bibitem[{\citenamefont{Nozawa et~al.}(1990)\citenamefont{Nozawa, Blankleider,
  and Lee}}]{Nozawa:1989pu}
\bibinfo{author}{\bibfnamefont{S.}~\bibnamefont{Nozawa}},
  \bibinfo{author}{\bibfnamefont{B.}~\bibnamefont{Blankleider}},
  \bibnamefont{and} \bibinfo{author}{\bibfnamefont{T.~S.~H.}
  \bibnamefont{Lee}}, \bibinfo{journal}{Nucl. Phys. A}
  \textbf{\bibinfo{volume}{513}}, \bibinfo{pages}{459} (\bibinfo{year}{1990}).

\bibitem[{\citenamefont{Lee and Pearce}(1991)}]{Lee:1991pp}
\bibinfo{author}{\bibfnamefont{T.~S.~H.} \bibnamefont{Lee}} \bibnamefont{and}
  \bibinfo{author}{\bibfnamefont{B.~C.} \bibnamefont{Pearce}},
  \bibinfo{journal}{Nucl. Phys. A} \textbf{\bibinfo{volume}{530}},
  \bibinfo{pages}{532} (\bibinfo{year}{1991}).

\bibitem[{\citenamefont{Davidson et~al.}(1991)\citenamefont{Davidson,
  Mukhopadhyay, and Wittman}}]{Davidson:1991xz}
\bibinfo{author}{\bibfnamefont{R.}~\bibnamefont{Davidson}},
  \bibinfo{author}{\bibfnamefont{N.}~\bibnamefont{Mukhopadhyay}},
  \bibnamefont{and} \bibinfo{author}{\bibfnamefont{R.}~\bibnamefont{Wittman}},
  \bibinfo{journal}{Phys. Rev. D} \textbf{\bibinfo{volume}{43}},
  \bibinfo{pages}{71} (\bibinfo{year}{1991}).

\bibitem[{\citenamefont{Garcilazo and Moya~de Guerra}(1993)}]{Garcilazo:1993av}
\bibinfo{author}{\bibfnamefont{H.}~\bibnamefont{Garcilazo}} \bibnamefont{and}
  \bibinfo{author}{\bibfnamefont{E.}~\bibnamefont{Moya~de Guerra}},
  \bibinfo{journal}{Nucl. Phys. A} \textbf{\bibinfo{volume}{562}},
  \bibinfo{pages}{521} (\bibinfo{year}{1993}).

\bibitem[{\citenamefont{Vanderhaeghen et~al.}(1995)\citenamefont{Vanderhaeghen,
  Heyde, Ryckebusch, and Waroquier}}]{Vanderhaeghen:1995fe}
\bibinfo{author}{\bibfnamefont{M.}~\bibnamefont{Vanderhaeghen}},
  \bibinfo{author}{\bibfnamefont{K.}~\bibnamefont{Heyde}},
  \bibinfo{author}{\bibfnamefont{J.}~\bibnamefont{Ryckebusch}},
  \bibnamefont{and}
  \bibinfo{author}{\bibfnamefont{M.}~\bibnamefont{Waroquier}},
  \bibinfo{journal}{Nucl. Phys. A} \textbf{\bibinfo{volume}{595}},
  \bibinfo{pages}{219} (\bibinfo{year}{1995}).

\bibitem[{\citenamefont{Pascalutsa and Scholten}(1995)}]{Pascalutsa:1995vx}
\bibinfo{author}{\bibfnamefont{V.}~\bibnamefont{Pascalutsa}} \bibnamefont{and}
  \bibinfo{author}{\bibfnamefont{O.}~\bibnamefont{Scholten}},
  \bibinfo{journal}{Nucl. Phys. A} \textbf{\bibinfo{volume}{591}},
  \bibinfo{pages}{658} (\bibinfo{year}{1995}).

\bibitem[{\citenamefont{David et~al.}(1996)\citenamefont{David, Fayard, Lamot,
  and Saghai}}]{David:1995pi}
\bibinfo{author}{\bibfnamefont{J.}~\bibnamefont{David}},
  \bibinfo{author}{\bibfnamefont{C.}~\bibnamefont{Fayard}},
  \bibinfo{author}{\bibfnamefont{G.}~\bibnamefont{Lamot}}, \bibnamefont{and}
  \bibinfo{author}{\bibfnamefont{B.}~\bibnamefont{Saghai}},
  \bibinfo{journal}{Phys. Rev. C} \textbf{\bibinfo{volume}{53}},
  \bibinfo{pages}{2613} (\bibinfo{year}{1996}).

\bibitem[{\citenamefont{Scholten et~al.}(1996)\citenamefont{Scholten, Korchin,
  Pascalutsa, and Van~Neck}}]{Scholten:1996mw}
\bibinfo{author}{\bibfnamefont{O.}~\bibnamefont{Scholten}},
  \bibinfo{author}{\bibfnamefont{A.}~\bibnamefont{Korchin}},
  \bibinfo{author}{\bibfnamefont{V.}~\bibnamefont{Pascalutsa}},
  \bibnamefont{and} \bibinfo{author}{\bibfnamefont{D.}~\bibnamefont{Van~Neck}},
  \bibinfo{journal}{Phys. Lett. B} \textbf{\bibinfo{volume}{384}},
  \bibinfo{pages}{13} (\bibinfo{year}{1996}), \eprint{nucl-th/9604014}.

\bibitem[{\citenamefont{Feuster and Mosel}(1997)}]{Feuster:1996ww}
\bibinfo{author}{\bibfnamefont{T.}~\bibnamefont{Feuster}} \bibnamefont{and}
  \bibinfo{author}{\bibfnamefont{U.}~\bibnamefont{Mosel}},
  \bibinfo{journal}{Nucl. Phys. A} \textbf{\bibinfo{volume}{612}},
  \bibinfo{pages}{375} (\bibinfo{year}{1997}), \eprint{nucl-th/9604026}.

\bibitem[{\citenamefont{Pascalutsa and Tjon}(1998)}]{Pascalutsa:1997md}
\bibinfo{author}{\bibfnamefont{V.}~\bibnamefont{Pascalutsa}} \bibnamefont{and}
  \bibinfo{author}{\bibfnamefont{J.~A.} \bibnamefont{Tjon}},
  \bibinfo{journal}{Nucl. Phys. A} \textbf{\bibinfo{volume}{631}},
  \bibinfo{pages}{534C} (\bibinfo{year}{1998}), \eprint{nucl-th/9709017}.

\bibitem[{\citenamefont{Korchin et~al.}(1998)\citenamefont{Korchin, Scholten,
  and Timmermans}}]{Korchin:1998ff}
\bibinfo{author}{\bibfnamefont{A.}~\bibnamefont{Korchin}},
  \bibinfo{author}{\bibfnamefont{O.}~\bibnamefont{Scholten}}, \bibnamefont{and}
  \bibinfo{author}{\bibfnamefont{R.}~\bibnamefont{Timmermans}},
  \bibinfo{journal}{Phys. Lett. B} \textbf{\bibinfo{volume}{438}},
  \bibinfo{pages}{1} (\bibinfo{year}{1998}), \eprint{nucl-th/9811042}.

\bibitem[{\citenamefont{Mota et~al.}(1999)\citenamefont{Mota, Garcilazo,
  Valcarce, and Fernandez}}]{Mota:1999qp}
\bibinfo{author}{\bibfnamefont{R.}~\bibnamefont{Mota}},
  \bibinfo{author}{\bibfnamefont{H.}~\bibnamefont{Garcilazo}},
  \bibinfo{author}{\bibfnamefont{A.}~\bibnamefont{Valcarce}}, \bibnamefont{and}
  \bibinfo{author}{\bibfnamefont{F.}~\bibnamefont{Fernandez}},
  \bibinfo{journal}{Phys. Rev. C} \textbf{\bibinfo{volume}{59}},
  \bibinfo{pages}{46} (\bibinfo{year}{1999}).

\bibitem[{\citenamefont{Lahiff and Afnan}(1999)}]{Lahiff:1999ur}
\bibinfo{author}{\bibfnamefont{A.}~\bibnamefont{Lahiff}} \bibnamefont{and}
  \bibinfo{author}{\bibfnamefont{I.}~\bibnamefont{Afnan}},
  \bibinfo{journal}{Phys. Rev. C} \textbf{\bibinfo{volume}{60}},
  \bibinfo{pages}{024608} (\bibinfo{year}{1999}), \eprint{nucl-th/9903058}.

\bibitem[{\citenamefont{Freedman et~al.}(1976)\citenamefont{Freedman, van
  Nieuwenhuizen, and Ferrara}}]{Freedman:1976xh}
\bibinfo{author}{\bibfnamefont{D.~Z.} \bibnamefont{Freedman}},
  \bibinfo{author}{\bibfnamefont{P.}~\bibnamefont{van Nieuwenhuizen}},
  \bibnamefont{and} \bibinfo{author}{\bibfnamefont{S.}~\bibnamefont{Ferrara}},
  \bibinfo{journal}{Phys. Rev. D} \textbf{\bibinfo{volume}{13}},
  \bibinfo{pages}{3214} (\bibinfo{year}{1976}).

\bibitem[{\citenamefont{Deser and Zumino}(1976)}]{Deser:1976eh}
\bibinfo{author}{\bibfnamefont{S.}~\bibnamefont{Deser}} \bibnamefont{and}
  \bibinfo{author}{\bibfnamefont{B.}~\bibnamefont{Zumino}},
  \bibinfo{journal}{Phys. Lett. B} \textbf{\bibinfo{volume}{62}},
  \bibinfo{pages}{335} (\bibinfo{year}{1976}).

\bibitem[{\citenamefont{Freedman and van
  Nieuwenhuizen}(1976)}]{Freedman:1976py}
\bibinfo{author}{\bibfnamefont{D.~Z.} \bibnamefont{Freedman}} \bibnamefont{and}
  \bibinfo{author}{\bibfnamefont{P.}~\bibnamefont{van Nieuwenhuizen}},
  \bibinfo{journal}{Phys. Rev. D} \textbf{\bibinfo{volume}{14}},
  \bibinfo{pages}{912} (\bibinfo{year}{1976}).

\bibitem[{\citenamefont{Van~Nieuwenhuizen}(1981)}]{VanNieuwenhuizen:1981ae}
\bibinfo{author}{\bibfnamefont{P.}~\bibnamefont{Van~Nieuwenhuizen}},
  \bibinfo{journal}{Phys. Rept.} \textbf{\bibinfo{volume}{68}},
  \bibinfo{pages}{189} (\bibinfo{year}{1981}).

\bibitem[{\citenamefont{Grisaru and Pendleton}(1977)}]{Grisaru:1977kk}
\bibinfo{author}{\bibfnamefont{M.~T.} \bibnamefont{Grisaru}} \bibnamefont{and}
  \bibinfo{author}{\bibfnamefont{H.}~\bibnamefont{Pendleton}},
  \bibinfo{journal}{Phys. Lett. B} \textbf{\bibinfo{volume}{67}},
  \bibinfo{pages}{323} (\bibinfo{year}{1977}).

\bibitem[{\citenamefont{Grisaru et~al.}(1977)\citenamefont{Grisaru, Pendleton,
  and van Nieuwenhuizen}}]{Grisaru:1976vm}
\bibinfo{author}{\bibfnamefont{M.~T.} \bibnamefont{Grisaru}},
  \bibinfo{author}{\bibfnamefont{H.}~\bibnamefont{Pendleton}},
  \bibnamefont{and} \bibinfo{author}{\bibfnamefont{P.}~\bibnamefont{van
  Nieuwenhuizen}}, \bibinfo{journal}{Phys. Rev. D}
  \textbf{\bibinfo{volume}{15}}, \bibinfo{pages}{996} (\bibinfo{year}{1977}).

\bibitem[{\citenamefont{Hagen and Singh}(1982)}]{Hagen:1982ez}
\bibinfo{author}{\bibfnamefont{C.}~\bibnamefont{Hagen}} \bibnamefont{and}
  \bibinfo{author}{\bibfnamefont{L.}~\bibnamefont{Singh}},
  \bibinfo{journal}{Phys. Rev. D} \textbf{\bibinfo{volume}{26}},
  \bibinfo{pages}{393} (\bibinfo{year}{1982}).

\bibitem[{\citenamefont{Pascalutsa}(1998)}]{Pascalutsa:1998pw}
\bibinfo{author}{\bibfnamefont{V.}~\bibnamefont{Pascalutsa}},
  \bibinfo{journal}{Phys. Rev. D} \textbf{\bibinfo{volume}{58}},
  \bibinfo{pages}{096002} (\bibinfo{year}{1998}), \eprint{hep-ph/9802288}.

\bibitem[{\citenamefont{Pascalutsa and Timmermans}(1999)}]{Pascalutsa:1999zz}
\bibinfo{author}{\bibfnamefont{V.}~\bibnamefont{Pascalutsa}} \bibnamefont{and}
  \bibinfo{author}{\bibfnamefont{R.}~\bibnamefont{Timmermans}},
  \bibinfo{journal}{Phys. Rev. C} \textbf{\bibinfo{volume}{60}},
  \bibinfo{pages}{042201} (\bibinfo{year}{1999}), \eprint{nucl-th/9905065}.

\bibitem[{\citenamefont{Pascalutsa}(2001)}]{Pascalutsa:2000kd}
\bibinfo{author}{\bibfnamefont{V.}~\bibnamefont{Pascalutsa}},
  \bibinfo{journal}{Phys. Lett. B} \textbf{\bibinfo{volume}{503}},
  \bibinfo{pages}{85} (\bibinfo{year}{2001}), \eprint{hep-ph/0008026}.

\bibitem[{\citenamefont{Badagnani et~al.}(2017)\citenamefont{Badagnani,
  Mariano, and Barbero}}]{Badagnani:2017una}
\bibinfo{author}{\bibfnamefont{D.}~\bibnamefont{Badagnani}},
  \bibinfo{author}{\bibfnamefont{A.}~\bibnamefont{Mariano}}, \bibnamefont{and}
  \bibinfo{author}{\bibfnamefont{C.}~\bibnamefont{Barbero}},
  \bibinfo{journal}{J. Phys. G} \textbf{\bibinfo{volume}{44}},
  \bibinfo{pages}{025001} (\bibinfo{year}{2017}).

\bibitem[{\citenamefont{Badagnani
  et~al.}(2015{\natexlab{a}})\citenamefont{Badagnani, Barbero, and
  Mariano}}]{Badagnani:2015rwj}
\bibinfo{author}{\bibfnamefont{D.}~\bibnamefont{Badagnani}},
  \bibinfo{author}{\bibfnamefont{C.}~\bibnamefont{Barbero}}, \bibnamefont{and}
  \bibinfo{author}{\bibfnamefont{A.}~\bibnamefont{Mariano}},
  \bibinfo{journal}{J. Phys. G} \textbf{\bibinfo{volume}{42}},
  \bibinfo{pages}{125001} (\bibinfo{year}{2015}{\natexlab{a}}).

\bibitem[{\citenamefont{Badagnani
  et~al.}(2015{\natexlab{b}})\citenamefont{Badagnani, Barbero, and
  Mariano}}]{Badagnani:2015pfa}
\bibinfo{author}{\bibfnamefont{D.}~\bibnamefont{Badagnani}},
  \bibinfo{author}{\bibfnamefont{C.}~\bibnamefont{Barbero}}, \bibnamefont{and}
  \bibinfo{author}{\bibfnamefont{A.}~\bibnamefont{Mariano}}
  (\bibinfo{year}{2015}{\natexlab{b}}), \eprint{1503.01612}.

\bibitem[{\citenamefont{Bernard et~al.}(1996)\citenamefont{Bernard, Kaiser, and
  Meissner}}]{Bernard:1994gm}
\bibinfo{author}{\bibfnamefont{V.}~\bibnamefont{Bernard}},
  \bibinfo{author}{\bibfnamefont{N.}~\bibnamefont{Kaiser}}, \bibnamefont{and}
  \bibinfo{author}{\bibfnamefont{U.-G.} \bibnamefont{Meissner}},
  \bibinfo{journal}{Z. Phys. C} \textbf{\bibinfo{volume}{70}},
  \bibinfo{pages}{483} (\bibinfo{year}{1996}), \eprint{hep-ph/9411287}.

\bibitem[{\citenamefont{Bernard et~al.}(1995)\citenamefont{Bernard, Kaiser, and
  Meissner}}]{Bernard:1995dp}
\bibinfo{author}{\bibfnamefont{V.}~\bibnamefont{Bernard}},
  \bibinfo{author}{\bibfnamefont{N.}~\bibnamefont{Kaiser}}, \bibnamefont{and}
  \bibinfo{author}{\bibfnamefont{U.-G.} \bibnamefont{Meissner}},
  \bibinfo{journal}{Int. J. Mod. Phys. E} \textbf{\bibinfo{volume}{4}},
  \bibinfo{pages}{193} (\bibinfo{year}{1995}), \eprint{hep-ph/9501384}.

\bibitem[{\citenamefont{Hemmert et~al.}(1997)\citenamefont{Hemmert, Holstein,
  and Kambor}}]{Hemmert:1996xg}
\bibinfo{author}{\bibfnamefont{T.~R.} \bibnamefont{Hemmert}},
  \bibinfo{author}{\bibfnamefont{B.~R.} \bibnamefont{Holstein}},
  \bibnamefont{and} \bibinfo{author}{\bibfnamefont{J.}~\bibnamefont{Kambor}},
  \bibinfo{journal}{Phys. Lett. B} \textbf{\bibinfo{volume}{395}},
  \bibinfo{pages}{89} (\bibinfo{year}{1997}), \eprint{hep-ph/9606456}.

\bibitem[{\citenamefont{Tang and Ellis}(1996)}]{Tang:1996sq}
\bibinfo{author}{\bibfnamefont{H.-B.} \bibnamefont{Tang}} \bibnamefont{and}
  \bibinfo{author}{\bibfnamefont{P.~J.} \bibnamefont{Ellis}},
  \bibinfo{journal}{Phys. Lett. B} \textbf{\bibinfo{volume}{387}},
  \bibinfo{pages}{9} (\bibinfo{year}{1996}), \eprint{hep-ph/9606432}.

\bibitem[{\citenamefont{Hemmert et~al.}(1998)\citenamefont{Hemmert, Holstein,
  and Kambor}}]{Hemmert:1997ye}
\bibinfo{author}{\bibfnamefont{T.~R.} \bibnamefont{Hemmert}},
  \bibinfo{author}{\bibfnamefont{B.~R.} \bibnamefont{Holstein}},
  \bibnamefont{and} \bibinfo{author}{\bibfnamefont{J.}~\bibnamefont{Kambor}},
  \bibinfo{journal}{J. Phys. G} \textbf{\bibinfo{volume}{24}},
  \bibinfo{pages}{1831} (\bibinfo{year}{1998}), \eprint{hep-ph/9712496}.

\bibitem[{\citenamefont{Fettes et~al.}(1998)\citenamefont{Fettes, Meissner, and
  Steininger}}]{Fettes:1998ud}
\bibinfo{author}{\bibfnamefont{N.}~\bibnamefont{Fettes}},
  \bibinfo{author}{\bibfnamefont{U.-G.} \bibnamefont{Meissner}},
  \bibnamefont{and}
  \bibinfo{author}{\bibfnamefont{S.}~\bibnamefont{Steininger}},
  \bibinfo{journal}{Nucl. Phys. A} \textbf{\bibinfo{volume}{640}},
  \bibinfo{pages}{199} (\bibinfo{year}{1998}), \eprint{hep-ph/9803266}.

\bibitem[{\citenamefont{Hacker et~al.}(2005)\citenamefont{Hacker, Wies,
  Gegelia, and Scherer}}]{Hacker:2005fh}
\bibinfo{author}{\bibfnamefont{C.}~\bibnamefont{Hacker}},
  \bibinfo{author}{\bibfnamefont{N.}~\bibnamefont{Wies}},
  \bibinfo{author}{\bibfnamefont{J.}~\bibnamefont{Gegelia}}, \bibnamefont{and}
  \bibinfo{author}{\bibfnamefont{S.}~\bibnamefont{Scherer}},
  \bibinfo{journal}{Phys. Rev. C} \textbf{\bibinfo{volume}{72}},
  \bibinfo{pages}{055203} (\bibinfo{year}{2005}), \eprint{hep-ph/0505043}.

\bibitem[{\citenamefont{Wies et~al.}(2006)\citenamefont{Wies, Gegelia, and
  Scherer}}]{Wies:2006rv}
\bibinfo{author}{\bibfnamefont{N.}~\bibnamefont{Wies}},
  \bibinfo{author}{\bibfnamefont{J.}~\bibnamefont{Gegelia}}, \bibnamefont{and}
  \bibinfo{author}{\bibfnamefont{S.}~\bibnamefont{Scherer}},
  \bibinfo{journal}{Phys. Rev. D} \textbf{\bibinfo{volume}{73}},
  \bibinfo{pages}{094012} (\bibinfo{year}{2006}), \eprint{hep-ph/0602073}.

\bibitem[{\citenamefont{Pascalutsa et~al.}(2007)\citenamefont{Pascalutsa,
  Vanderhaeghen, and Yang}}]{Pascalutsa:2006up}
\bibinfo{author}{\bibfnamefont{V.}~\bibnamefont{Pascalutsa}},
  \bibinfo{author}{\bibfnamefont{M.}~\bibnamefont{Vanderhaeghen}},
  \bibnamefont{and} \bibinfo{author}{\bibfnamefont{S.~N.} \bibnamefont{Yang}},
  \bibinfo{journal}{Phys. Rept.} \textbf{\bibinfo{volume}{437}},
  \bibinfo{pages}{125} (\bibinfo{year}{2007}), \eprint{hep-ph/0609004}.

\bibitem[{\citenamefont{Mai et~al.}(2012)\citenamefont{Mai, Bruns, and
  Meissner}}]{Mai:2012wy}
\bibinfo{author}{\bibfnamefont{M.}~\bibnamefont{Mai}},
  \bibinfo{author}{\bibfnamefont{P.~C.} \bibnamefont{Bruns}}, \bibnamefont{and}
  \bibinfo{author}{\bibfnamefont{U.-G.} \bibnamefont{Meissner}},
  \bibinfo{journal}{Phys. Rev. D} \textbf{\bibinfo{volume}{86}},
  \bibinfo{pages}{094033} (\bibinfo{year}{2012}), \eprint{1207.4923}.

\bibitem[{\citenamefont{Chen et~al.}(2013)\citenamefont{Chen, Yao, and
  Zheng}}]{Chen:2012nx}
\bibinfo{author}{\bibfnamefont{Y.-H.} \bibnamefont{Chen}},
  \bibinfo{author}{\bibfnamefont{D.-L.} \bibnamefont{Yao}}, \bibnamefont{and}
  \bibinfo{author}{\bibfnamefont{H.~Q.} \bibnamefont{Zheng}},
  \bibinfo{journal}{Phys. Rev. D} \textbf{\bibinfo{volume}{87}},
  \bibinfo{pages}{054019} (\bibinfo{year}{2013}), \eprint{1212.1893}.

\bibitem[{\citenamefont{Hilt et~al.}(2013)\citenamefont{Hilt, Scherer, and
  Tiator}}]{Hilt:2013uf}
\bibinfo{author}{\bibfnamefont{M.}~\bibnamefont{Hilt}},
  \bibinfo{author}{\bibfnamefont{S.}~\bibnamefont{Scherer}}, \bibnamefont{and}
  \bibinfo{author}{\bibfnamefont{L.}~\bibnamefont{Tiator}},
  \bibinfo{journal}{Phys. Rev. C} \textbf{\bibinfo{volume}{87}},
  \bibinfo{pages}{045204} (\bibinfo{year}{2013}), \eprint{1301.5576}.

\bibitem[{\citenamefont{Siemens et~al.}(2016)\citenamefont{Siemens, Bernard,
  Epelbaum, Gasparyan, Krebs, and Mei\ss{}ner}}]{Siemens:2016hdi}
\bibinfo{author}{\bibfnamefont{D.}~\bibnamefont{Siemens}},
  \bibinfo{author}{\bibfnamefont{V.}~\bibnamefont{Bernard}},
  \bibinfo{author}{\bibfnamefont{E.}~\bibnamefont{Epelbaum}},
  \bibinfo{author}{\bibfnamefont{A.}~\bibnamefont{Gasparyan}},
  \bibinfo{author}{\bibfnamefont{H.}~\bibnamefont{Krebs}}, \bibnamefont{and}
  \bibinfo{author}{\bibfnamefont{U.-G.} \bibnamefont{Mei\ss{}ner}},
  \bibinfo{journal}{Phys. Rev. C} \textbf{\bibinfo{volume}{94}},
  \bibinfo{pages}{014620} (\bibinfo{year}{2016}), \eprint{1602.02640}.

\bibitem[{\citenamefont{Yao et~al.}(2016)\citenamefont{Yao, Siemens, Bernard,
  Epelbaum, Gasparyan, Gegelia, Krebs, and Mei\ss{}ner}}]{Yao:2016vbz}
\bibinfo{author}{\bibfnamefont{D.-L.} \bibnamefont{Yao}},
  \bibinfo{author}{\bibfnamefont{D.}~\bibnamefont{Siemens}},
  \bibinfo{author}{\bibfnamefont{V.}~\bibnamefont{Bernard}},
  \bibinfo{author}{\bibfnamefont{E.}~\bibnamefont{Epelbaum}},
  \bibinfo{author}{\bibfnamefont{A.~M.} \bibnamefont{Gasparyan}},
  \bibinfo{author}{\bibfnamefont{J.}~\bibnamefont{Gegelia}},
  \bibinfo{author}{\bibfnamefont{H.}~\bibnamefont{Krebs}}, \bibnamefont{and}
  \bibinfo{author}{\bibfnamefont{U.-G.} \bibnamefont{Mei\ss{}ner}},
  \bibinfo{journal}{JHEP} \textbf{\bibinfo{volume}{05}}, \bibinfo{pages}{038}
  (\bibinfo{year}{2016}), \eprint{1603.03638}.

\bibitem[{\citenamefont{Hiller~Blin et~al.}(2016)\citenamefont{Hiller~Blin,
  Ledwig, and Vicente~Vacas}}]{Blin:2016itn}
\bibinfo{author}{\bibfnamefont{A.~N.} \bibnamefont{Hiller~Blin}},
  \bibinfo{author}{\bibfnamefont{T.}~\bibnamefont{Ledwig}}, \bibnamefont{and}
  \bibinfo{author}{\bibfnamefont{M.~J.} \bibnamefont{Vicente~Vacas}},
  \bibinfo{journal}{Phys. Rev. D} \textbf{\bibinfo{volume}{93}},
  \bibinfo{pages}{094018} (\bibinfo{year}{2016}), \eprint{1602.08967}.

\bibitem[{\citenamefont{Guerrero~Navarro
  et~al.}(2019)\citenamefont{Guerrero~Navarro, Vicente~Vacas, Blin, and
  Yao}}]{Navarro:2019iqj}
\bibinfo{author}{\bibfnamefont{G.~H.} \bibnamefont{Guerrero~Navarro}},
  \bibinfo{author}{\bibfnamefont{M.~J.} \bibnamefont{Vicente~Vacas}},
  \bibinfo{author}{\bibfnamefont{A.~N.~H.} \bibnamefont{Blin}},
  \bibnamefont{and} \bibinfo{author}{\bibfnamefont{D.-L.} \bibnamefont{Yao}},
  \bibinfo{journal}{Phys. Rev. D} \textbf{\bibinfo{volume}{100}},
  \bibinfo{pages}{094021} (\bibinfo{year}{2019}), \eprint{1908.00890}.

\bibitem[{\citenamefont{Mart et~al.}(2015)\citenamefont{Mart, Clymton, and
  Arifi}}]{Mart:2015jof}
\bibinfo{author}{\bibfnamefont{T.}~\bibnamefont{Mart}},
  \bibinfo{author}{\bibfnamefont{S.}~\bibnamefont{Clymton}}, \bibnamefont{and}
  \bibinfo{author}{\bibfnamefont{A.~J.} \bibnamefont{Arifi}},
  \bibinfo{journal}{Phys. Rev. D} \textbf{\bibinfo{volume}{92}},
  \bibinfo{pages}{094019} (\bibinfo{year}{2015}).

\bibitem[{\citenamefont{Mart}(2019)}]{Mart:2019mtq}
\bibinfo{author}{\bibfnamefont{T.}~\bibnamefont{Mart}}, \bibinfo{journal}{Phys.
  Rev. D} \textbf{\bibinfo{volume}{100}}, \bibinfo{pages}{056008}
  (\bibinfo{year}{2019}), \eprint{1909.02696}.

\bibitem[{\citenamefont{Clymton and Mart}(2021)}]{Clymton:2021wof}
\bibinfo{author}{\bibfnamefont{S.}~\bibnamefont{Clymton}} \bibnamefont{and}
  \bibinfo{author}{\bibfnamefont{T.}~\bibnamefont{Mart}}
  (\bibinfo{year}{2021}), \eprint{2104.10333}.

\bibitem[{\citenamefont{Anisovich et~al.}(2010)\citenamefont{Anisovich, Klempt,
  Nikonov, Matveev, Sarantsev, and Thoma}}]{Anisovich:2009zy}
\bibinfo{author}{\bibfnamefont{A.~V.} \bibnamefont{Anisovich}},
  \bibinfo{author}{\bibfnamefont{E.}~\bibnamefont{Klempt}},
  \bibinfo{author}{\bibfnamefont{V.~A.} \bibnamefont{Nikonov}},
  \bibinfo{author}{\bibfnamefont{M.~A.} \bibnamefont{Matveev}},
  \bibinfo{author}{\bibfnamefont{A.~V.} \bibnamefont{Sarantsev}},
  \bibnamefont{and} \bibinfo{author}{\bibfnamefont{U.}~\bibnamefont{Thoma}},
  \bibinfo{journal}{Eur. Phys. J. A} \textbf{\bibinfo{volume}{44}},
  \bibinfo{pages}{203} (\bibinfo{year}{2010}), \eprint{0911.5277}.

\bibitem[{\citenamefont{Anisovich et~al.}(2012)\citenamefont{Anisovich, Beck,
  Klempt, Nikonov, Sarantsev, and Thoma}}]{Anisovich:2011fc}
\bibinfo{author}{\bibfnamefont{A.~V.} \bibnamefont{Anisovich}},
  \bibinfo{author}{\bibfnamefont{R.}~\bibnamefont{Beck}},
  \bibinfo{author}{\bibfnamefont{E.}~\bibnamefont{Klempt}},
  \bibinfo{author}{\bibfnamefont{V.~A.} \bibnamefont{Nikonov}},
  \bibinfo{author}{\bibfnamefont{A.~V.} \bibnamefont{Sarantsev}},
  \bibnamefont{and} \bibinfo{author}{\bibfnamefont{U.}~\bibnamefont{Thoma}},
  \bibinfo{journal}{Eur. Phys. J. A} \textbf{\bibinfo{volume}{48}},
  \bibinfo{pages}{15} (\bibinfo{year}{2012}), \eprint{1112.4937}.

\bibitem[{\citenamefont{Kamano et~al.}(2013)\citenamefont{Kamano, Nakamura,
  Lee, and Sato}}]{Kamano:2013iva}
\bibinfo{author}{\bibfnamefont{H.}~\bibnamefont{Kamano}},
  \bibinfo{author}{\bibfnamefont{S.~X.} \bibnamefont{Nakamura}},
  \bibinfo{author}{\bibfnamefont{T.~S.~H.} \bibnamefont{Lee}},
  \bibnamefont{and} \bibinfo{author}{\bibfnamefont{T.}~\bibnamefont{Sato}},
  \bibinfo{journal}{Phys. Rev. C} \textbf{\bibinfo{volume}{88}},
  \bibinfo{pages}{035209} (\bibinfo{year}{2013}), \eprint{1305.4351}.

\bibitem[{\citenamefont{R\"onchen et~al.}(2014)\citenamefont{R\"onchen,
  D\"oring, Huang, Haberzettl, Haidenbauer, Hanhart, Krewald, Mei\ss{}ner, and
  Nakayama}}]{Ronchen:2014cna}
\bibinfo{author}{\bibfnamefont{D.}~\bibnamefont{R\"onchen}},
  \bibinfo{author}{\bibfnamefont{M.}~\bibnamefont{D\"oring}},
  \bibinfo{author}{\bibfnamefont{F.}~\bibnamefont{Huang}},
  \bibinfo{author}{\bibfnamefont{H.}~\bibnamefont{Haberzettl}},
  \bibinfo{author}{\bibfnamefont{J.}~\bibnamefont{Haidenbauer}},
  \bibinfo{author}{\bibfnamefont{C.}~\bibnamefont{Hanhart}},
  \bibinfo{author}{\bibfnamefont{S.}~\bibnamefont{Krewald}},
  \bibinfo{author}{\bibfnamefont{U.~G.} \bibnamefont{Mei\ss{}ner}},
  \bibnamefont{and} \bibinfo{author}{\bibfnamefont{K.}~\bibnamefont{Nakayama}},
  \bibinfo{journal}{Eur. Phys. J. A} \textbf{\bibinfo{volume}{50}},
  \bibinfo{pages}{101} (\bibinfo{year}{2014}), \bibinfo{note}{[Erratum:
  Eur.Phys.J.A 51, 63 (2015)]}, \eprint{1401.0634}.

\bibitem[{\citenamefont{Sekihara}(2017)}]{Sekihara:2016xnq}
\bibinfo{author}{\bibfnamefont{T.}~\bibnamefont{Sekihara}},
  \bibinfo{journal}{Phys. Rev. C} \textbf{\bibinfo{volume}{95}},
  \bibinfo{pages}{025206} (\bibinfo{year}{2017}), \eprint{1609.09496}.

\bibitem[{\citenamefont{Sekihara}(2021)}]{Sekihara:2021eah}
\bibinfo{author}{\bibfnamefont{T.}~\bibnamefont{Sekihara}}
  (\bibinfo{year}{2021}), \eprint{2104.01962}.

\bibitem[{\citenamefont{Pearce and Jennings}(1991)}]{Pearce:1990uj}
\bibinfo{author}{\bibfnamefont{B.~C.} \bibnamefont{Pearce}} \bibnamefont{and}
  \bibinfo{author}{\bibfnamefont{B.~K.} \bibnamefont{Jennings}},
  \bibinfo{journal}{Nucl. Phys. A} \textbf{\bibinfo{volume}{528}},
  \bibinfo{pages}{655} (\bibinfo{year}{1991}).

\bibitem[{\citenamefont{Gross and Surya}(1993)}]{Gross:1992tj}
\bibinfo{author}{\bibfnamefont{F.}~\bibnamefont{Gross}} \bibnamefont{and}
  \bibinfo{author}{\bibfnamefont{Y.}~\bibnamefont{Surya}},
  \bibinfo{journal}{Phys. Rev. C} \textbf{\bibinfo{volume}{47}},
  \bibinfo{pages}{703} (\bibinfo{year}{1993}).

\bibitem[{\citenamefont{Sato and Lee}(1996)}]{Sato:1996gk}
\bibinfo{author}{\bibfnamefont{T.}~\bibnamefont{Sato}} \bibnamefont{and}
  \bibinfo{author}{\bibfnamefont{T.-. S.~H.} \bibnamefont{Lee}},
  \bibinfo{journal}{Phys. Rev. C} \textbf{\bibinfo{volume}{54}},
  \bibinfo{pages}{2660} (\bibinfo{year}{1996}), \eprint{nucl-th/9606009}.

\bibitem[{\citenamefont{Feuster and Mosel}(1998)}]{Feuster:1997pq}
\bibinfo{author}{\bibfnamefont{T.}~\bibnamefont{Feuster}} \bibnamefont{and}
  \bibinfo{author}{\bibfnamefont{U.}~\bibnamefont{Mosel}},
  \bibinfo{journal}{Phys. Rev. C} \textbf{\bibinfo{volume}{58}},
  \bibinfo{pages}{457} (\bibinfo{year}{1998}), \eprint{nucl-th/9708051}.

\bibitem[{\citenamefont{Feuster and Mosel}(1999)}]{Feuster:1998cj}
\bibinfo{author}{\bibfnamefont{T.}~\bibnamefont{Feuster}} \bibnamefont{and}
  \bibinfo{author}{\bibfnamefont{U.}~\bibnamefont{Mosel}},
  \bibinfo{journal}{Phys. Rev. C} \textbf{\bibinfo{volume}{59}},
  \bibinfo{pages}{460} (\bibinfo{year}{1999}), \eprint{nucl-th/9803057}.

\bibitem[{\citenamefont{Pascalutsa and Tjon}(2000)}]{Pascalutsa:2000bs}
\bibinfo{author}{\bibfnamefont{V.}~\bibnamefont{Pascalutsa}} \bibnamefont{and}
  \bibinfo{author}{\bibfnamefont{J.~A.} \bibnamefont{Tjon}},
  \bibinfo{journal}{Phys. Rev. C} \textbf{\bibinfo{volume}{61}},
  \bibinfo{pages}{054003} (\bibinfo{year}{2000}), \eprint{nucl-th/0003050}.

\bibitem[{\citenamefont{Thapa et~al.}(2021)\citenamefont{Thapa, Sinha, Li, and
  Sedrakian}}]{Thapa:2021kfo}
\bibinfo{author}{\bibfnamefont{V.~B.} \bibnamefont{Thapa}},
  \bibinfo{author}{\bibfnamefont{M.}~\bibnamefont{Sinha}},
  \bibinfo{author}{\bibfnamefont{J.~J.} \bibnamefont{Li}}, \bibnamefont{and}
  \bibinfo{author}{\bibfnamefont{A.}~\bibnamefont{Sedrakian}},
  \bibinfo{journal}{Phys. Rev. D} \textbf{\bibinfo{volume}{103}},
  \bibinfo{pages}{063004} (\bibinfo{year}{2021}), \eprint{2102.08787}.

\bibitem[{\citenamefont{Raduta}(2021)}]{Raduta:2021xiz}
\bibinfo{author}{\bibfnamefont{A.~R.} \bibnamefont{Raduta}},
  \bibinfo{journal}{Phys. Lett. B} \textbf{\bibinfo{volume}{814}},
  \bibinfo{pages}{136070} (\bibinfo{year}{2021}), \eprint{2101.03718}.

\bibitem[{\citenamefont{Li et~al.}(2020)\citenamefont{Li, Sedrakian, and
  Weber}}]{Li:2020ias}
\bibinfo{author}{\bibfnamefont{J.~J.} \bibnamefont{Li}},
  \bibinfo{author}{\bibfnamefont{A.}~\bibnamefont{Sedrakian}},
  \bibnamefont{and} \bibinfo{author}{\bibfnamefont{F.}~\bibnamefont{Weber}},
  \bibinfo{journal}{Phys. Lett. B} \textbf{\bibinfo{volume}{810}},
  \bibinfo{pages}{135812} (\bibinfo{year}{2020}), \eprint{2010.02901}.

\bibitem[{\citenamefont{Silvi et~al.}(2021)}]{Silvi:2021uya}
\bibinfo{author}{\bibfnamefont{G.}~\bibnamefont{Silvi}} \bibnamefont{et~al.}
  (\bibinfo{year}{2021}), \eprint{2101.00689}.

\bibitem[{\citenamefont{Weinberg}(1964)}]{Weinberg:1964cn}
\bibinfo{author}{\bibfnamefont{S.}~\bibnamefont{Weinberg}},
  \bibinfo{journal}{Phys. Rev.} \textbf{\bibinfo{volume}{133}},
  \bibinfo{pages}{B1318} (\bibinfo{year}{1964}).

\bibitem[{\citenamefont{Delgado-Acosta and
  Kirchbach}(2013)}]{Delgado-Acosta:2013kva}
\bibinfo{author}{\bibfnamefont{E.~G.} \bibnamefont{Delgado-Acosta}}
  \bibnamefont{and} \bibinfo{author}{\bibfnamefont{M.}~\bibnamefont{Kirchbach}}
  (\bibinfo{year}{2013}), \eprint{1312.5811}.

\bibitem[{\citenamefont{Delgado~Acosta
  et~al.}(2015)\citenamefont{Delgado~Acosta, Banda~Guzman, and
  Kirchbach}}]{DelgadoAcosta:2015ypa}
\bibinfo{author}{\bibfnamefont{E.}~\bibnamefont{Delgado~Acosta}},
  \bibinfo{author}{\bibfnamefont{V.}~\bibnamefont{Banda~Guzman}},
  \bibnamefont{and}
  \bibinfo{author}{\bibfnamefont{M.}~\bibnamefont{Kirchbach}},
  \bibinfo{journal}{Eur. Phys. J. A} \textbf{\bibinfo{volume}{51}},
  \bibinfo{pages}{35} (\bibinfo{year}{2015}), \eprint{1503.07230}.

\bibitem[{\citenamefont{Mart et~al.}(2019)\citenamefont{Mart, Kristiano, and
  Clymton}}]{Mart:2019jtb}
\bibinfo{author}{\bibfnamefont{T.}~\bibnamefont{Mart}},
  \bibinfo{author}{\bibfnamefont{J.}~\bibnamefont{Kristiano}},
  \bibnamefont{and} \bibinfo{author}{\bibfnamefont{S.}~\bibnamefont{Clymton}},
  \bibinfo{journal}{Phys. Rev.} \textbf{\bibinfo{volume}{C100}},
  \bibinfo{pages}{035207} (\bibinfo{year}{2019}), \eprint{1909.04282}.

\bibitem[{\citenamefont{Criado et~al.}(2020)\citenamefont{Criado, Koivunen,
  Raidal, and Veerm\"ae}}]{Criado:2020jkp}
\bibinfo{author}{\bibfnamefont{J.~C.} \bibnamefont{Criado}},
  \bibinfo{author}{\bibfnamefont{N.}~\bibnamefont{Koivunen}},
  \bibinfo{author}{\bibfnamefont{M.}~\bibnamefont{Raidal}}, \bibnamefont{and}
  \bibinfo{author}{\bibfnamefont{H.}~\bibnamefont{Veerm\"ae}},
  \bibinfo{journal}{Phys. Rev. D} \textbf{\bibinfo{volume}{102}},
  \bibinfo{pages}{125031} (\bibinfo{year}{2020}), \eprint{2010.02224}.

\bibitem[{\citenamefont{Criado et~al.}(2021{\natexlab{a}})\citenamefont{Criado,
  Djouadi, Koivunen, Raidal, and Veerm\"ae}}]{Criado:2021itq}
\bibinfo{author}{\bibfnamefont{J.~C.} \bibnamefont{Criado}},
  \bibinfo{author}{\bibfnamefont{A.}~\bibnamefont{Djouadi}},
  \bibinfo{author}{\bibfnamefont{N.}~\bibnamefont{Koivunen}},
  \bibinfo{author}{\bibfnamefont{M.}~\bibnamefont{Raidal}}, \bibnamefont{and}
  \bibinfo{author}{\bibfnamefont{H.}~\bibnamefont{Veerm\"ae}}
  (\bibinfo{year}{2021}{\natexlab{a}}), \eprint{2102.13652}.

\bibitem[{\citenamefont{Criado et~al.}(2021{\natexlab{b}})\citenamefont{Criado,
  Djouadi, Koivunen, M\"u\"ursepp, Raidal, and Veerm\"ae}}]{Criado:2021qpd}
\bibinfo{author}{\bibfnamefont{J.~C.} \bibnamefont{Criado}},
  \bibinfo{author}{\bibfnamefont{A.}~\bibnamefont{Djouadi}},
  \bibinfo{author}{\bibfnamefont{N.}~\bibnamefont{Koivunen}},
  \bibinfo{author}{\bibfnamefont{K.}~\bibnamefont{M\"u\"ursepp}},
  \bibinfo{author}{\bibfnamefont{M.}~\bibnamefont{Raidal}}, \bibnamefont{and}
  \bibinfo{author}{\bibfnamefont{H.}~\bibnamefont{Veerm\"ae}}
  (\bibinfo{year}{2021}{\natexlab{b}}), \eprint{2104.03231}.

\bibitem[{\citenamefont{Haberzettl et~al.}(1998)\citenamefont{Haberzettl,
  Bennhold, Mart, and Feuster}}]{Haberzettl:1998aqi}
\bibinfo{author}{\bibfnamefont{H.}~\bibnamefont{Haberzettl}},
  \bibinfo{author}{\bibfnamefont{C.}~\bibnamefont{Bennhold}},
  \bibinfo{author}{\bibfnamefont{T.}~\bibnamefont{Mart}}, \bibnamefont{and}
  \bibinfo{author}{\bibfnamefont{T.}~\bibnamefont{Feuster}},
  \bibinfo{journal}{Phys. Rev. C} \textbf{\bibinfo{volume}{58}},
  \bibinfo{pages}{R40} (\bibinfo{year}{1998}), \eprint{nucl-th/9804051}.

\bibitem[{\citenamefont{Klempt and Richard}(2010)}]{Klempt:2009pi}
\bibinfo{author}{\bibfnamefont{E.}~\bibnamefont{Klempt}} \bibnamefont{and}
  \bibinfo{author}{\bibfnamefont{J.-M.} \bibnamefont{Richard}},
  \bibinfo{journal}{Rev. Mod. Phys.} \textbf{\bibinfo{volume}{82}},
  \bibinfo{pages}{1095} (\bibinfo{year}{2010}), \eprint{0901.2055}.

\bibitem[{\citenamefont{Epelbaum et~al.}(2009)\citenamefont{Epelbaum, Hammer,
  and Meissner}}]{Epelbaum:2008ga}
\bibinfo{author}{\bibfnamefont{E.}~\bibnamefont{Epelbaum}},
  \bibinfo{author}{\bibfnamefont{H.-W.} \bibnamefont{Hammer}},
  \bibnamefont{and} \bibinfo{author}{\bibfnamefont{U.-G.}
  \bibnamefont{Meissner}}, \bibinfo{journal}{Rev. Mod. Phys.}
  \textbf{\bibinfo{volume}{81}}, \bibinfo{pages}{1773} (\bibinfo{year}{2009}),
  \eprint{0811.1338}.

\bibitem[{\citenamefont{Weinberg}(1979)}]{Weinberg:1978kz}
\bibinfo{author}{\bibfnamefont{S.}~\bibnamefont{Weinberg}},
  \bibinfo{journal}{Physica A} \textbf{\bibinfo{volume}{96}},
  \bibinfo{pages}{327} (\bibinfo{year}{1979}).

\bibitem[{\citenamefont{Gasser and Leutwyler}(1984)}]{Gasser:1983yg}
\bibinfo{author}{\bibfnamefont{J.}~\bibnamefont{Gasser}} \bibnamefont{and}
  \bibinfo{author}{\bibfnamefont{H.}~\bibnamefont{Leutwyler}},
  \bibinfo{journal}{Annals Phys.} \textbf{\bibinfo{volume}{158}},
  \bibinfo{pages}{142} (\bibinfo{year}{1984}).

\bibitem[{\citenamefont{Gasser and Leutwyler}(1985)}]{Gasser:1984gg}
\bibinfo{author}{\bibfnamefont{J.}~\bibnamefont{Gasser}} \bibnamefont{and}
  \bibinfo{author}{\bibfnamefont{H.}~\bibnamefont{Leutwyler}},
  \bibinfo{journal}{Nucl. Phys. B} \textbf{\bibinfo{volume}{250}},
  \bibinfo{pages}{465} (\bibinfo{year}{1985}).

\bibitem[{\citenamefont{Vrancx et~al.}(2011)\citenamefont{Vrancx, De~Cruz,
  Ryckebusch, and Vancraeyveld}}]{Vrancx:2011qv}
\bibinfo{author}{\bibfnamefont{T.}~\bibnamefont{Vrancx}},
  \bibinfo{author}{\bibfnamefont{L.}~\bibnamefont{De~Cruz}},
  \bibinfo{author}{\bibfnamefont{J.}~\bibnamefont{Ryckebusch}},
  \bibnamefont{and}
  \bibinfo{author}{\bibfnamefont{P.}~\bibnamefont{Vancraeyveld}},
  \bibinfo{journal}{Phys. Rev. C} \textbf{\bibinfo{volume}{84}},
  \bibinfo{pages}{045201} (\bibinfo{year}{2011}), \eprint{1105.2688}.

\bibitem[{\citenamefont{Dreiner et~al.}(2010)\citenamefont{Dreiner, Haber, and
  Martin}}]{Dreiner:2008tw}
\bibinfo{author}{\bibfnamefont{H.~K.} \bibnamefont{Dreiner}},
  \bibinfo{author}{\bibfnamefont{H.~E.} \bibnamefont{Haber}}, \bibnamefont{and}
  \bibinfo{author}{\bibfnamefont{S.~P.} \bibnamefont{Martin}},
  \bibinfo{journal}{Phys. Rept.} \textbf{\bibinfo{volume}{494}},
  \bibinfo{pages}{1} (\bibinfo{year}{2010}), \eprint{0812.1594}.

\bibitem[{\citenamefont{Borodulin et~al.}(2017)\citenamefont{Borodulin,
  Rogalyov, and Slabospitskii}}]{Borodulin:2017pwh}
\bibinfo{author}{\bibfnamefont{V.~I.} \bibnamefont{Borodulin}},
  \bibinfo{author}{\bibfnamefont{R.~N.} \bibnamefont{Rogalyov}},
  \bibnamefont{and} \bibinfo{author}{\bibfnamefont{S.~R.}
  \bibnamefont{Slabospitskii}} (\bibinfo{year}{2017}), \eprint{1702.08246}.

\end{thebibliography}

\end{document}